\documentclass[aps,pre,twocolumn,amssymb,amsfonts,floatfix,showpacs,superscriptaddress]{revtex4}

\usepackage{graphicx}
\usepackage{dcolumn}
\usepackage{bm}
\usepackage{mathbbol}

\usepackage{amsmath,amssymb,amsthm}
\usepackage{times}
\usepackage{graphicx}
\usepackage{bm}
\usepackage{color}
\usepackage{multi row}

\begin{document}
\title{Local quenches with global effects in interacting quantum systems}

\author{E. J. Torres-Herrera}
\affiliation{Department of Physics, Yeshiva University, New York, New York 10016, USA}
\author{Lea F. Santos}
\affiliation{Department of Physics, Yeshiva University, New York, New York 10016, USA}

\begin{abstract}
We study one-dimensional lattices of interacting spins-$\frac{1}{2}$ and show that the effects of quenching the amplitude of a local magnetic field applied to a single site of the lattice can be comparable to the effects of a global perturbation  applied instantaneously to the entire system. Both quenches take the system to the chaotic domain, the energy distribution of the initial states approaches a Breit-Wigner shape, the fidelity (Loschmidt echo) decays exponentially, and thermalization becomes viable.
\end{abstract}


\maketitle

\section{Introduction}

In the broad field of nonequilibrium quantum physics, the unitary evolution of isolated systems after an instantaneous perturbation (quench) has become a prominent subject~\cite{Cazalilla2011,Polkovnikov2011RMP}. The enthusiasm is in part due to the development of computational methods to study strongly correlated quantum systems, such as density matrix renormalization group \cite{White2004,Schollwock2013} and numerical linked-cluster expansions \cite{Rigol2006,RigolARXIV}, and to ongoing experiments with nuclear magnetic resonance~\cite{Cappellaro2007,Ramanathan2011,KaurARXIV} and with cold atoms in optical lattices~\cite{Greiner2002,Kinoshita06,hofferberth07,Bloch2008,Trotzky2008,Simon2011,Trotzky2012,Fukuhara2013}. In the latter case, the high level of control and quasi-isolation allow for the experimental analysis of coherent evolutions for very long-times. 

The studies of quench dynamics often distinguish global from local perturbations. What is referred to as one or the other presupposes a choice of basis representation. Local quenches in space have been addressed in~\cite{Fukuhara2013,Calabrese2007,Eisler2007,Eisler2008,Calabrese2009,Diez2010,Stephan2011,Divakaran2011,Smacchia2012,Ganahl2012,Nozaki2013,AsplundARXIV,AlbaARXIV}. The experiment in~\cite{Fukuhara2013} analyzes the quantum dynamics of an excitation created by flipping a single spin in the middle of a Heisenberg chain. The case where two semi-infinite lines are joined at their endpoints and subsequently evolved as a single infinite system has received special attention, because of the possibility of achieving exact results using conformal field theory~\cite{Calabrese2007}. Other theoretical studies include the quench of a local magnetic field applied to few sites of a Heisenberg spin-$\frac{1}{2}$ chain~\cite{Diez2010} and a time dependent local quench of the transverse field in the Ising chain~\cite{Smacchia2012}. The general expectation is that the effects of local quenches be more limited than those of global quenches~\cite{Stephan2011}. 

Here, we show that for the systems and initial states considered, local and global quenches in space may in fact lead to equivalent outcomes. The local quench that we investigate is similar to the one treated in~\cite{Diez2010}. The system is initially in an eigenstate of an initial Hamiltonian $\widehat{H}_I$ that describes an integrable clean anisotropic Heisenberg spin-$\frac{1}{2}$ chain with only nearest-neighbor (NN) couplings, the so-called $XXZ$ model. We quench the amplitude of a magnetic field applied to a {\em single site} from zero to a finite value, creating an on-site defect (impurity). The consequences of such, at first sight, minor perturbation are drastic. The interplay between the defect and the interactions of the $XXZ$ model takes the system into the chaotic domain~\cite{Santos2004,Barisic2009,Santos2011}. This reminds one of the Sinai's billiards, where the square table becomes chaotic due to a circular obstacle of arbitrarily small radius placed at its center~\cite{Sinai1970,Berry1981}. We demonstrate that the local quench of the magnetic field is comparable to a global quench where couplings between next-nearest neighbors (NNN) are suddenly included in the clean $XXZ$ model. The frustrated Heisenberg spin-$\frac{1}{2}$ Hamiltonian that emerges with the addition of NNN terms is also chaotic.

Depending on the amplitude of the on-site static field and on the ratio between NNN and NN couplings, both final Hamiltonians $\widehat{H}_F$, for the impurity and frustrated chains, look alike when written in the basis coinciding with the eigenstates of $\widehat{H}_I$. From this correspondence, a list of similarities between both types of quenches follows. For initial states away from the edges of the spectrum, the fidelity (Loschmidt echo), which measures the probability of finding the initial state in time, decays exponentially. The Shannon (information) entropy in the basis of $\widehat{H}_I$ increases linearly in time. The comparison between infinite time averages and microcanonical averages for local and nonlocal few-body observables and different system sizes indicates the viability of thermalization.

Our local and global quenches coincide in the limit of intermediate perturbation. In this scenario, the energy distributions of the initial states have a Breit-Wigner (Lorentzian) shape~\cite{ZelevinskyRep1996,Jacquod2001,Flambaum2001a,Flambaum2001b,Emerson2002,Weinstein2003,Izrailev2006,Santos2012PRL,Santos2012PRE}, which leads to the exponential decay of the fidelity mentioned above. By further increasing the ratio between NNN and NN couplings, the frustrated system eventually reaches the strong perturbation regime. At this point, the initial states approach the maximum possible level of delocalization available to a system with two-body interactions, their energy distributions achieving a Gaussian form and the fidelity decay becoming Gaussian~\cite{Flambaum1994,Frazier1996,ZelevinskyRep1996,Flambaum2000,Izrailev2006,
Santos2012PRL,Santos2012PRE,Torres2014PRA,Torres2014NJP,GenwayPRL2010}. Our studies suggest that this limit cannot be reached by the local quench.

This paper is organized as follows. Section~\ref{Sec:model} presents the models and quenches studied. Section~\ref{Sec:chaos} compares the Hamiltonian matrices for the impurity and NNN models, as well as their eigenvalues and eigenstates. The results for the fidelity decay and the evolution of the Shannon entropy are given in Sec.~\ref{Sec:RelaxationProcess}. The viability of thermalization is discussed and illustrated in Sec.~\ref{Sec:Thermalization}. Concluding remarks are presented in Sec.~\ref{Sec:Summary}.

\section{System Models and Quenches}
\label{Sec:model}

We investigate a one-dimensional spin-$\frac{1}{2}$ system with $L$ sites and open boundary conditions. Spin-$\frac{1}{2}$ systems are the prototype of realistic quantum systems with interactions. They describe real magnetic compounds~\cite{Sologubenko2000PRB,Hess2007,Hlubek2010}, crystals of fluorapatite~\cite{Cappellaro2007,Ramanathan2011,KaurARXIV}, and can also be simulated with optical lattices~\cite{Trotzky2008,Simon2011,Trotzky2012,Fukuhara2013}. The model we consider contains only two-body interactions and is described by the following Hamiltonian
\begin{equation}\label{eq:Ham}
\widehat{H} = \varepsilon J \widehat{S}_{1}^z  + d J \widehat{S}_{\lfloor L/2 \rfloor }^z + \widehat{H}_{\text{NN}}+ \lambda \widehat{H}_{\text{NNN}},
\end{equation}
where
\begin{eqnarray}
&& \widehat{H}_{\text{NN}} = J\sum_{i=1}^{L-1} \left( 
\widehat{S}_i^x \widehat{S}_{i+1}^x + \widehat{S}_i^y \widehat{S}_{i+1}^y +
\Delta \widehat{S}_i^z \widehat{S}_{i+1}^z \right),  
 \\
&& \widehat{H}_{\text{NNN}} = J\sum_{i=1}^{L-2} \left(\widehat{S}_i^x \widehat{S}_{i+2}^x + \widehat{S}_i^y \widehat{S}_{i+2}^y
+ \Delta \widehat{S}_i^z \widehat{S}_{i+2}^z  \right). 
\end{eqnarray}
Above, $\hbar =1$ and $\widehat{S}^{x,y,z}_i =\widehat{\sigma}^{x,y,z}_i/2$ are spin operators acting on site $i$, $\widehat{\sigma}^{x,y,z}_i$ being the Pauli matrices. $\widehat{S}_i^x \widehat{S}_{i+1}^x + \widehat{S}_i^y \widehat{S}_{i+1}^y$ $(\widehat{S}_i^x \widehat{S}_{i+2}^x + \widehat{S}_i^y \widehat{S}_{i+2}^y)$ is the flip-flop term and $\widehat{S}_i^z \widehat{S}_{i+1}^z (\widehat{S}_i^z \widehat{S}_{i+2}^z)$ is the Ising interaction between NN (NNN) spins. $J$ is the exchange coupling constant, $\Delta$ is the anisotropy parameter, and $\lambda$ refers to the ratio between NNN and NN couplings. These three parameters are assumed positive, thus favoring antiferromagnetic order. 

In this work we compare the properties of the spin-$\frac{1}{2}$ model above in two chaotic limits. For the signatures of quantum chaos to be revealed, the spectrum needs to be separated according to symmetry sectors. If energies from different subspaces are mixed, such signatures (discussed in Sec.III.B) may be concealed even when the system is chaotic~\cite{Santos2009JMP,Gubin2012}. At the same time, good statistics requires large subspaces. Thus, a good strategy is to prevent the emergence of too many symmetries.  In our case, Hamiltonian (\ref{eq:Ham}) conserves total spin in the $z$-direction, $\widehat{{\cal{S}}}^z=\sum_i\widehat{S}_i^z$, for any chosen parameters, but other symmetries  are avoided as follows.

(i) Translational symmetry is avoided by choosing open boundary conditions. 

(ii) By working outside the ${\cal{S}}^z=0$ sector, we avoid spin reversal symmetry, that is invariance under a $\pi$ rotation around the $x$ axis. We choose to deal with the subspace that has $L/3$ up spins, which still has a large dimension, ${\cal D}=L!/[(2L/3)!(L/3)!]$, for the largest system sizes that can be handled with exact diagonalization. We consider $L=12,15$, and $18$.

(iii) When the Ising interaction is present, we use $\Delta=0.48$. This is sufficiently away from the midpoint $\Delta=1/2$, where the system develops additional nontrivial symmetries (see references in \cite{Zangara2013}), and it prevents conservation of total spin, $S^2=(\sum_{i=1}^L \vec{S}_i)^2$, which happens at $\Delta=1$.

(iv) To avoid reflection symmetry, we add a small impurity of amplitude $\varepsilon J $ on the first site of the chain. It can be generated by applying a local static magnetic field in the $z$-direction. We fix $\varepsilon = 0.1 $. This defect does not break the integrability of the system~\cite{Alcaraz1987}. Notice that it also weakly breaks the symmetries described in (i), (ii), and (iii). 

A second local magnetic field may be placed close to the middle of the chain, on site $\lfloor L/2 \rfloor$, leading to the Zeeman splitting $d J$. Depending on the values of the parameters $\Delta, d$, and $\lambda$, the chain may be integrable or chaotic:

{\em Integrable XX model}: $\Delta, \lambda =0$. This Hamiltonian is solved with the Jordan-Wigner transformation, which maps the system onto a model of noninteracting spinless fermions~\cite{Jordan1928}. 

{\em Integrable $XXZ$ model}: $\Delta\neq 0$ and $d, \lambda=0$ . This model is solved by means of the Bethe ansatz~\cite{Bethe1931}. 

Notice that we call XX and $XXZ$ the clean models ($d=0$) with NN couplings only ($\lambda=0$).    

{\em Chaotic impurity model}: $\Delta, d \neq 0$ and $\lambda=0$. The addition of a single impurity close to the middle of the chain in the presence of NN couplings can bring the system into the chaotic domain~\cite{Santos2004,Barisic2009,Santos2011}. The onset of chaos is caused by the interplay between the Ising interaction and the impurity. In contrast, the addition of $d$ to the XX model does not affect its integrability. In the interacting system, chaoticity requires $d \lesssim 1$. If the defect becomes too large, it splits the system in two independent and integrable chains. This motivates our choice of $d=0.9$ in most of the work.

{\em Chaotic NNN model}: $\Delta, \lambda \neq 0 $ and $d=0$. The addition of couplings between second neighbors breaks integrability~\cite{Kudo2004,Kudo2005,Gubin2012}. There are different combinations of parameters that can lead to chaos~\cite{Santos2011,Gubin2012}.  We consider the complete case where both NN and NNN flip-flop and Ising terms are present. 

As discussed in Sec.~III, the two chaotic models above show very similar properties when $\lambda$ is relatively small ({\em e.g} when $\lambda \sim 0.4$), but differences become noticeable when $\lambda \rightarrow 1$.                                                                                                                                                                                                                                                                                                                                                                                                                                                                                                                                                                                                                                                                                                                                                                                                                                                                                                                                                            

\subsection{Local and global quenches}
\label{Sec:quench}

Our system starts in an excited eigenstate of the initial Hamiltonian corresponding to the integrable $XXZ$ model with a small defect on site 1:
\begin{equation}
\widehat{H}_I = \varepsilon J \widehat{S}_{1}^z + \widehat{H}_{\text{NN}}
\end{equation}
We analyze the short time dynamics and infinite time averages after the following two instantaneous perturbations.
\begin{itemize}
\item \emph{Local quench}. The perturbation is localized on a single site:  $d_\text{I}= 0 \rightarrow d_\text{F} \neq 0$. $\widehat{H}_I$ is quenched to the chaotic impurity model with NN couplings only,
\begin{equation}
\widehat{H}_F^{\text{local}} = \widehat{H}_I + d_F J \widehat{S}_{\lfloor L/2 \rfloor }^z .
\end{equation}
\end{itemize}
\begin{itemize}
\item \emph{Global quench}. The perturbation affects simultaneously all sites in the chain: $\lambda_\text{I}=0\rightarrow \lambda_\text{F} \neq 0$. $\widehat{H}_I$ is quenched to the chaotic Hamiltonian with NNN couplings,
\end{itemize}
\begin{equation}
\widehat{H}_F^{\text{global}} = \widehat{H}_I +  \lambda_F \widehat{H}_{\text{NNN}} .
\end{equation}

\section{Impurity vs NNN couplings}
\label{Sec:chaos}

For appropriate values of $d_F$ and $\lambda_F$, the structure of the Hamiltonian matrices of the impurity and NNN models and the results for the signatures of chaos associated with their eigenvalues and eigenstates become comparable.

\subsection{Structure of the Hamiltonian matrices}

\begin{figure}[htb]
\centering
\includegraphics*[width=0.48\textwidth]{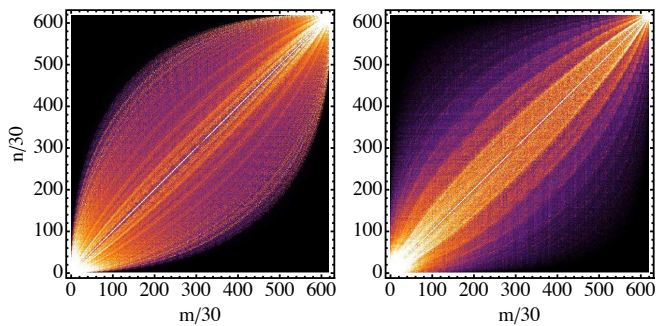}
\caption{(Color online) Absolute values of the elements of the impurity (left) and NNN (right) Hamiltonian matrices written in the basis corresponding to the eigenstates of $\widehat{H}_I$;  $\Delta=0.48$, $d_F=0.9$, $\lambda_F =0.44$, $L=18$, ${\cal{D}}=18564$. The basis is ordered in energy. Lighter color indicates larger values.}
\label{fig:matrix}	
\end{figure}

A natural choice is to write the final Hamiltonians in the basis corresponding to the eigenstates $|n\rangle$ of $\widehat{H}_I$. These states constitute the mean-field basis. We denote the $n^{th}$ eigenstate of $\widehat{H}_I$ by $|n\rangle$.  In this basis, the structure of the impurity and NNN Hamiltonian matrices can be very similar, as seen in Fig.~\ref{fig:matrix}. The diagonal elements, $\langle n|\widehat{H}_F |n \rangle \equiv H_{n,n}$, are large. They are ordered in energy, from low to high values. The off-diagonal elements, $\langle n|\widehat{H}_F |m \rangle \equiv H_{n,m}$, slowly fade away as the distance $|n-m|$ from the diagonal increases. The decay of the off-diagonal elements is typical of systems with two-(few-)body interactions and is in evident contrast with full random matrices~\cite{Brody1981}. 

\begin{figure}[htb]
\centering
\includegraphics*[width=0.48\textwidth]{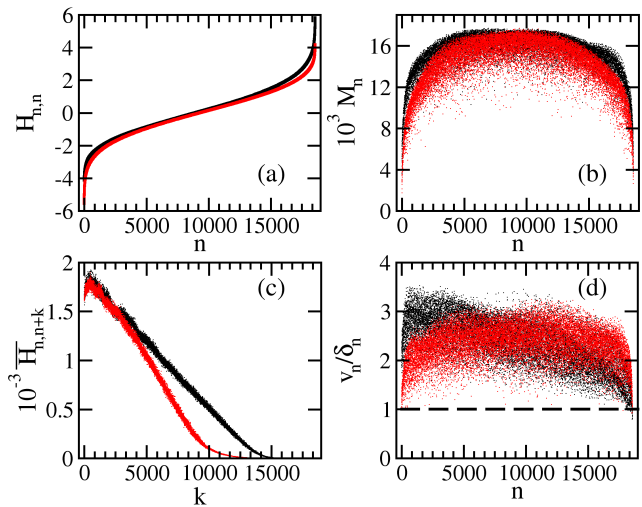}
\caption{(Color online)  Details of the Hamiltonian matrices of the impurity (light points, $d_F=0.9$) and NNN (dark points, $\lambda_F=0.44$) models written in the eigenstates of $\widehat{H}_I$; $\Delta=0.48$, $L=18$. Diagonal elements (a).  Connectivity (b). Averages of the absolute values of the off-diagonal elements vs the distance $k$ from the diagonal (c).  Ratio of the average coupling strength $v_n$ to the mean level spacing $\delta_n$ between directly coupled states in each line $n$ (d).}
\label{fig:matrixDetails}	
\end{figure}

The details of the matrices are better captured by Fig.~\ref{fig:matrixDetails}. Figure~\ref{fig:matrixDetails}(a) shows the values of the diagonal elements, which are very close for both models. Figure~\ref{fig:matrixDetails}(b) presents the values of the connectivity $M_n$ of each line $n$. The connectivity is the number of basis vectors directly coupled with each state $|n\rangle$, that is the number of nonzero $H_{n,m}$ for $n\neq m$. $M_n$ is comparable for both models. It shows a smooth behavior with $n$ (or equivalently with $H_{n,n}$). It is large in the middle of the spectrum, where the majority of the basis vectors are coupled, and it decreases at the edges~\cite{noteXXZ}. 

Notice that to count how many off-diagonal elements are nonzero, we use a threshold below which the elements are discarded. This is done because of our numerical procedure. Initially, $\widehat{H}_F$ is written in the natural site basis corresponding to product states of up and down spins. Subsequently, this basis is transformed into the states $|n\rangle$, which results in the appearance of many tiny off-diagonal elements not associated with any real coupling.  We use as threshold the variance of the absolute value of all off-diagonal elements.

In Fig.~\ref{fig:matrixDetails}(c), we show the averages of the absolute values of the off-diagonal elements,  
\begin{equation}
\overline{H}_{n,n+k}  =\frac{\sum_{n=1}^{{\cal D}-k} | H_{n,n+k} |}{{\cal D}-k},
\end{equation}
versus the distance $k$ from the diagonal. They are significantly smaller than the diagonal elements and decay with $k$. The absence of an abrupt drop implies that both Hamiltonians have long range, although finite interactions in the basis $|n\rangle$. Both kinds of perturbations to $\widehat{H}_I$ must therefore have nonlocal effects on the initial state. 

The similarity between $H_{n,n}$ and $M_n$ holds for both models when $d_F, \lambda_F \lesssim 1$, but differences are visible in the values of the off-diagonal elements. For parameters in the vicinity of those chosen in Fig.~\ref{fig:matrixDetails}(c), $\overline{H}_{n,n+k}$ is comparable for both systems when $k$ is small, but the decay is slower for the NNN model. Moreover, by increasing $\Delta$ ($d_F$ cannot be much increased, since we are already at the verge of splitting the chain), we can only slightly increase $\overline{H}_{n,n+k}$ for the impurity model and are unable to reach the large values achieved with $\lambda_F \rightarrow 1$ (not shown). In this limit, the NNN model can therefore lead to stronger mixing of the basis vectors than the defect case.

To get an idea of how effective the off-diagonal elements are, we compare their average strength
\begin{equation}
v_n = \frac{\sum_{m\neq n} |H_{n,m}|}{M_{n}}
\end{equation}
with the mean level spacing $\delta_n$ between directly coupled states. The latter is computed as
\begin{equation}
\delta_n=\frac{(H_{m,m})^{max}_n - (H_{m,m})^{min}_n }{M_{n}},
\end{equation}
where $(H_{m,m})^{max}_n$ [$(H_{m,m})^{min}_n$] is the largest (smallest) diagonal element where $H_{n,m} \neq0$. As seen in Fig.~\ref{fig:matrixDetails}(d), $v_n/\delta_n$ is similar for both models for the parameters considered. The ratio can be significantly increased by increasing $\lambda_F$, but it is hardly affected by larger combinations of $\Delta$ and $d_F$ (not shown). We can therefore distinguish between two limits: the intermediate perturbation regime, where $v_n/\delta_n \gtrsim 1$, and the strong perturbation regime, where $v_n/\delta_n$ can reach values significantly greater than 1. The second is only achieved by the NNN model.

\subsection{Eigenvalues and eigenstates}

In finite nonintegrable quantum systems, the appearance of properties that clearly indicate the onset of quantum chaos depends on the size of the perturbation. To determine how the system approaches the chaotic limit as the perturbation increases, we analyze the level spacing distribution and the level number variance~\cite{Guhr1998}. Both quantities require unfolding the spectrum of each symmetry sector separately. This procedure consists of locally rescaling the energies, so that the mean level density of the new sequence of energies is equal to one~\cite{Guhr1998,Gubin2012}. We also discard some few (10\% is our arbitrary choice) eigenvalues at the edges of the spectrum, where the fluctuations are large.

Quantum levels of integrable systems are not prohibited from crossing and the distribution of level spacings $s$ is typically Poissonian,
\begin{equation}
{\cal{P}}_{ P}(s) = \exp(-s).
\end{equation}
In chaotic systems, there is level repulsion and the level spacing distribution is given by the Wigner-Dyson distribution, as predicted by random matrix theory. Ensembles of full random matrices with time reversal invariance, the so-called Gaussian orthogonal ensembles (GOEs), lead to
\begin{equation}
{\cal{P}}_{\text{WD}}(s) = \frac{\pi s}{2} \exp \left( -\frac{\pi s^2}{4} \right).
\end{equation}
Despite the absence of randomness and the existence of only two-body interactions in the systems studied here, in the chaotic domain their ${\cal{P}}(s)$  is also given by the above Wigner-Dyson distribution. 

\begin{figure}[htb]
\centering
\includegraphics*[width=0.48\textwidth]{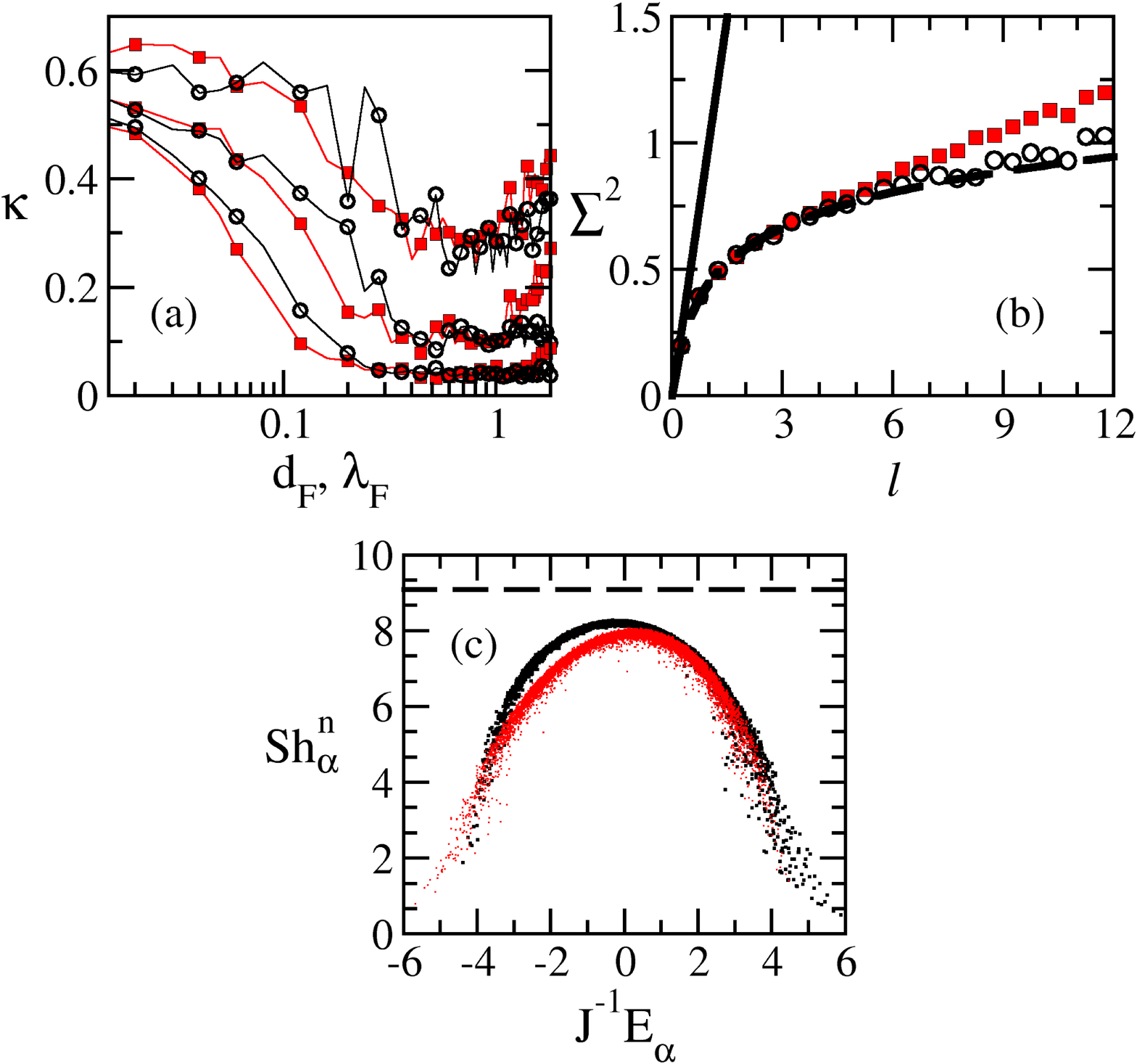}
\caption{(Color online) Indicator  $\kappa$ of the integrable-chaos crossover vs the perturbation strength (a), level number variance (b), and Shannon entropy for all eigenstates in the basis of $\widehat{H}_I$ (c) for the impurity (filled squares, light color) and the NNN (empty circles, dark color) models; $\Delta=0.48$. Panel (a):  $L=12,15$, and 18 from top to bottom. Panels (b) and (c): $L=18$, $d_F=0.9$, $\lambda_F=0.44$. Solid line: Poisson spectrum (b); dashed line: GOE result (b,c).}
\label{fig:Ps}	
\end{figure}

In finite systems, if the perturbation is not sufficiently large to reach ${\cal{P}}_\text{WD}(s)$, the level spacing distribution has an intermediate shape between Poisson and Wigner-Dyson.
To quantify the crossover from integrability to chaos, we show in Fig.~\ref{fig:Ps}(a) the level spacing indicator $\kappa$ defined as \cite{Santos2010PRE}
\begin{equation}
\kappa \equiv \frac{\sum_i[{\cal{P}}(s_i)-{\cal{P}}_{WD}(s_i)]}{\sum_i {\cal{P}}_{WD}(s_i)},
\label{eta}
\end{equation}
where the sums run over the whole spectrum. $\kappa$ is large close to the integrable domain and it approaches zero in the chaotic regime. This indicator 
is comparable to the quantity $\eta$ introduced in Ref.\ \cite{Jacquod1997} or the parameter $\beta$ used in the fitting of ${\cal{P}}(s)$ with the Brody distribution~\cite{Brody1981}. 

As the perturbations $d_F$ and $\lambda_F$ increase, both models become chaotic and show similar values of $\kappa$ for the same system sizes. If the perturbation is further increased well above 1, the systems eventually reach another integrable point. Notice also that as $L$ increases, the value of the perturbation leading to small $\kappa$ decreases. The onset of chaos in the thermodynamic limit might be achieved with an infinitesimally small integrability breaking term~\cite{Santos2010PRE}.

The level number variance  $\Sigma^2(l)$ quantifies long-range correlations~\cite{Guhr1998}. It measures the deviation of the staircase function from the best fit straight line. It is defined as
\begin{equation}
\Sigma^2(l) \equiv \langle N(l,g)^2 \rangle - \langle N(l,g)\rangle^2,
\end{equation}
where $N(l,g)$ gives the number of states in the interval $[g,g+l]$ and $\langle . \rangle $ represents the average over different initial values of $g$. For a Poisson distribution, $\Sigma^2(l)=l$, and for GOEs, $\Sigma^2(l)=2[\ln (2\pi l) + \gamma +1 -\pi^2/8 ]/\pi^2$, where $\gamma$ is the Euler constant. Level repulsion leads to rather rigid spectra and fluctuations are less significant than in regular systems. As shown in Fig.~\ref{fig:Ps}(b), the level number variances  for the impurity and NNN models are similar. They are also close to the GOE result. This proximity can be improved by changing the parameters, the best results being associated with the NNN Hamiltonian.

We also study the structure of the eigenstates $|\psi_{\alpha} \rangle$ of $\widehat{H}_F$ in the basis $|n \rangle $, 
\begin{equation}
|\psi_{\alpha} \rangle=\sum_{n} C_{\alpha}^{n} |n\rangle,
\end{equation}
via the Shannon (information) entropy,
\begin{equation}
\mbox{Sh}_{\alpha}^n \equiv -\sum_n  |C_{\alpha}^n|^2 \ln |C_{\alpha}^n|^2,
\label{entropyS}
\end{equation}
This delocalization measure determines the degree of complexity of the eigenstates. Complete delocalization occurs for full random matrices, where the amplitudes $C_{\alpha}^n$ are independent random variables. For GOEs, the average over the ensemble leads to  $\mbox{Sh}_{\text{GOE}} \sim \ln(0.48 {\cal D})$ \cite{Izrailev1990,ZelevinskyRep1996}. For the realistic systems considered here, where the Hamiltonian is sparse and banded, the mixing of the basis vectors is incomplete and $\mbox{Sh}_{\alpha}^n<\mbox{Sh}_{\text{GOE}} $.

Figure~\ref{fig:Ps}(c) compares the Shannon entropy for the defect and NNN models for all eigenstates. The results are very similar for the parameters considered. They reflect the structure of the matrices in Fig.~\ref{fig:matrixDetails}: $\mbox{Sh}_{\alpha}^n$ is large close to the middle of the spectrum, where $| H_{n,n} |$ is smaller and $M_n$ is larger, and it decreases as we approach the edges of the spectrum. This behavior mirrors also the density of states, which is Gaussian for systems with few-body interactions~\cite{Brody1981}. By increasing $\Delta$,  hardly any change is noticed on the values of the Shannon entropy for the impurity model, but by increasing $\lambda_F$, significantly larger values can be reached for the NNN model. This connects again with the notion of intermediate and strong perturbation regimes discussed in the description of Fig.~\ref{fig:matrixDetails}(d).

\section{Relaxation Dynamics}
\label{Sec:RelaxationProcess}

The correspondence between the static properties of the impurity and NNN models suggest that, starting from the same initial state, the dynamics of both systems should also be similar. This is confirmed in the following with the analysis of the fidelity decay and the evolution of the Shannon entropy.

\subsection{Local density of states and fidelity}
\label{Sec:LDOS}

The initial state  $|\Psi(0)\rangle = |\text{ini}\rangle$ evolves unitarily according to the eigenvalues $E_{\alpha}$ and eigenstates $|\psi_{\alpha}\rangle$ of $\widehat{H}_F$, as
\begin{equation}
|\Psi(t)\rangle=e^{-i\widehat{H}_\text{F}t}|\Psi(0)\rangle=\sum_{\alpha} C_{\alpha}^{\text{ini}}  e^{-iE_\alpha t}|\psi_\alpha\rangle .
\label{eq:instate}
\end{equation} 
The energy of $|\text{ini}\rangle$ projected on the final Hamiltonian is
\begin{equation}
E_{\text{ini}} = \langle \text{ini} |\widehat{H}_F | \text{ini} \rangle = \sum_{\alpha} |C_{\alpha}^{\text{ini}}|^2 E_{\alpha} .
\label{Eini}
\end{equation}
The eigenstate of $\widehat{H}_I$ that we select to be the initial state is the one for which $E_{\text{ini}}$ is closest to the energy
\begin{equation}
E_T=\frac{\sum_\alpha E_\alpha e^{-E_\alpha/k_B T}}{\sum_\alpha e^{-E_\alpha/k_B T}},
\end{equation}
fixed by a chosen temperature $T$. Above, $k_B$ is Boltzmann constant and it is set to 1.

The distribution $P^{\text{ini}}_{\alpha}$ of the components $|C_{\alpha}^{\text{ini}}|^2$ in the eigenvalues $E_{\alpha}$ is the so-called local density of states (LDOS) or strength function~\cite{Flambaum2000}. It corresponds to the energy distribution of the initial state and it gives information about the lifetime of the initial state. In particular, it determines how the fidelity decays in time.

The fidelity gives the probability of finding the system still in the initial state after time $t$. It corresponds to the overlap between $| \text{ini}\rangle $ and $ | \Psi(t) \rangle$~\cite{Gorin2006,Jacquod2009},
\begin{equation}
F(t) \equiv  |\langle \text{ini} | \Psi(t) \rangle |^2 = \left|\sum_{\alpha} |C_{\alpha}^{\text{ini}} |^2 e^{-i E_{\alpha} t}  \right|^2 .
\label{eq:fidelity}
\end{equation}
and is therefore equivalent to the Fourier transform in energy of the components $|C_{\alpha}^{\text{ini}}|^2$.

\begin{figure}[htb]
\centering
\includegraphics*[width=0.48\textwidth]{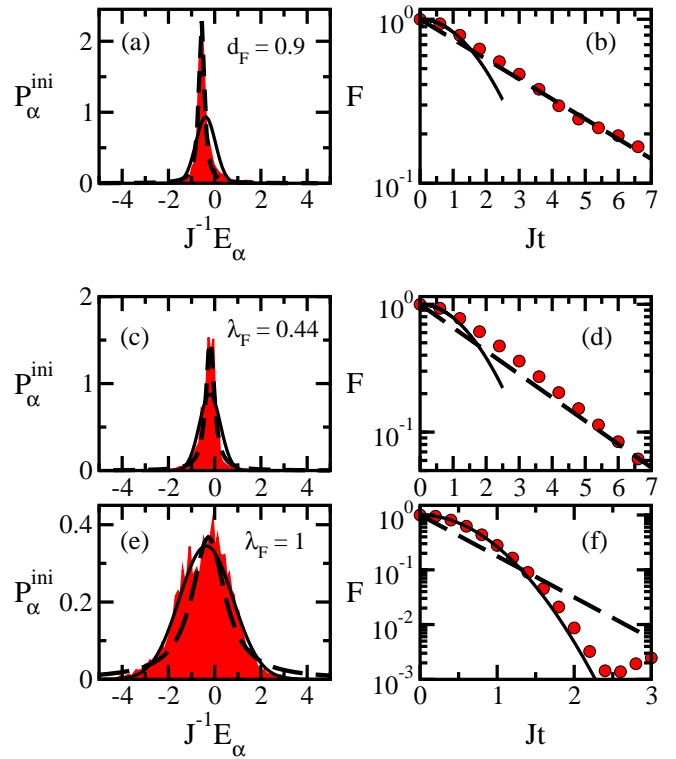}
\caption{(Color online) Local density of states (left) and fidelity decay (right) for the impurity (a), (b) and NNN (c), (d), (e), (f) models; $\Delta=0.48$, $L=18$. The values of the perturbation are indicated.  The initial state is an eigenstate of $\widehat{H}_I$ with $T=7J^{-1}$. Dashed line: Breit-Wigner fit (left) and corresponding exponential decay (right);
solid line: energy shell (left) and corresponding Gaussian decay (right); $\Gamma_{\text{ini}}=0.28$ and $\sigma_{\text{ini}}=0.42$ (a); $\Gamma_{\text{ini}}=0.42$ and $\sigma_{\text{ini}}=0.45$ (c); $\Gamma_{\text{ini}}=1.73$ and $\sigma_{\text{ini}}=1.16$ (e).}
\label{fig:LDOSandFidelity}	
\end{figure}

In the limit of intermediate perturbation and for $E_{\text{ini}}$ away from the edges of the spectrum, $P^{\text{ini}}_{\alpha}$ approaches a Breit-Wigner form delineated by ~\cite{ZelevinskyRep1996,Jacquod2001,Flambaum2001a,Flambaum2001b,Emerson2002,Weinstein2003,Izrailev2006,Santos2012PRL,Santos2012PRE}
\begin{equation}
P^{\text{ini}}_{BW}(E) = \frac{1}{2\pi} \frac{\Gamma_{\text{ini}} }{(E_{\text{ini}} - E)^2 +
 \Gamma_{\text{ini}}^2 /4},
\label{BW_SF}
\end{equation}
as shown in Figs.~\ref{fig:LDOSandFidelity}(a) and \ref{fig:LDOSandFidelity}(c) for the impurity and NNN models, respectively. A Lorentzian LDOS leads to the exponential decay of the fidelity,
\begin{equation}
F_{\text{BW}}(t) = \left| \int_{-\infty}^{\infty} P^{\text{ini}}_{BW}(E) e^{-i E t} dE
\right|^2
= e^{- \Gamma_{\text{ini}} t},
\label{Fbw}
\end{equation}
which is approximately the behavior seen in Figs.~\ref{fig:LDOSandFidelity}(b) and \ref{fig:LDOSandFidelity}(d) \cite{noteAditi}. We notice that the Breit-Wigner form of the LDOS is robust for the NNN model, that is the shape is maintained for small variations in the values of $\Delta, \lambda_F$ and $E_{\text{ini}}$, but for the impurity model fluctuations are observed. In this case, the shape is sometimes not well defined, with more than a main peak or a secondary bump deforming the Breit-Wigner.

For the NNN model, in the limit of strong perturbation, the LDOS of initial states away from the edges of the spectrum approaches a Gaussian shape, as shown in Fig.~\ref{fig:LDOSandFidelity}(e). The distribution is limited by the energy shell~\cite{Flambaum1994,Frazier1996,ZelevinskyRep1996,Flambaum2000,Flambaum2001a,
Flambaum2001b,Izrailev2006,Santos2012PRL,Santos2012PRE,Torres2014PRA,Torres2014NJP}, which is the Gaussian
\begin{equation}
P^{\text{ini}}_{G}(E) = \frac{1}{ \sqrt{ 2 \pi \sigma^2_{\text{ini}} }  } \exp \left[ -\frac{(E-E_{\text{ini}})^2}{ 2 \sigma^2_{\text{ini}} }   \right]
\label{Gauss_SF}
\end{equation}
with width
\begin{equation}
\sigma_{\text{ini}} = \sqrt{\sum_{\alpha} |C_{\alpha}^{\text{ini}} |^2 (E_{\alpha} - E_{\text{ini}})^2}
=\sqrt{\sum_{n \neq \text{ini}} |\langle n |\widehat{H}_F | \text{ini}\rangle |^2 }.
\label{deltaE}
\end{equation}
The shell determines the maximum possible spreading of $|\text{ini} \rangle$ in a system with two-body interactions. The Gaussian distribution induces the Gaussian decay of the fidelity,
\begin{equation}
F_{\text{G}}(t) = \left| \int_{-\infty}^{\infty} P^{\text{ini}}_{G}(E) e^{-i E t} dE
\right|^2
= e^{- \sigma_{\text{ini}}^2  t^2}.
\label{eq:Fgauss}
\end{equation}
as illustrated in Fig.\ref{fig:LDOSandFidelity}(f). The impurity model cannot reach this extreme scenario. Its LDOS does not go beyond the Breit-Wigner form and its fidelity decay is thus restricted to the exponential behavior.

We notice, however, that the fidelity decay of the impurity model can be significantly accelerated, being dictated by $\cos^2(d_F t/2)$, if one considers a very large defect. In the case where $d_F \gg 1$, $P^{\text{ini}}_{\alpha}$ is bimodal and the distance between the two peaks controls the initial dynamics.

\subsection{Evolution of the Shannon entropy}

\begin{figure}[htb]
\centering
\includegraphics*[width=0.48\textwidth]{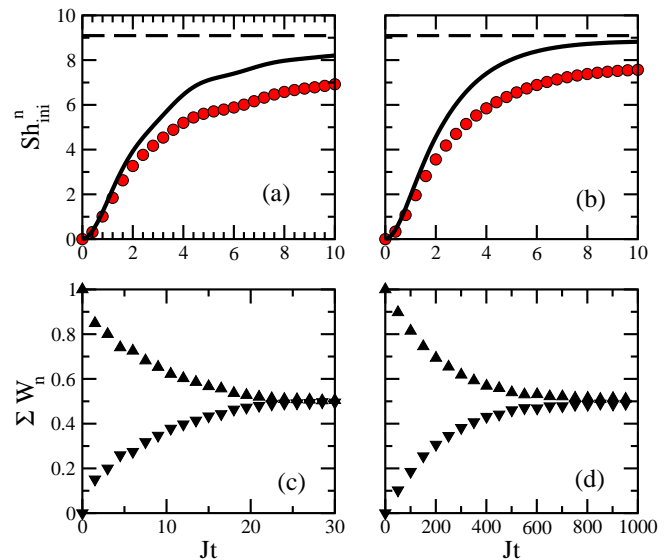}
\caption{(Color online) Shannon entropy $\text{Sh}_{\text{ini}}^n$ vs time (top) and sum of the probabilities $W_n$ separated for odd and even states $|n\rangle$ (bottom) for the impurity (a), (c) and NNN (b), (d) models. Circles: numerical results for $\text{Sh}_{\text{ini}}^n$; solid lines: semi-analytical Eq.~(\ref{eq:Sanalytical}); up triangles: $\sum_n W_n$ for states with the same parity as $|\text{ini}\rangle$; down triangles: states with the opposite parity. The initial state is an eigenstate of $\widehat{H}_I$ with $T=7 J^{-1}$; $\Delta=0.48$, $d_F=0.9$, $\lambda_F=0.44$, $L=18$.
$N_{pc}$ was obtained from the average in the interval $Jt \in [1000,2000]$. }
\label{fig:ShannonTime}	
\end{figure}

In Figs.~\ref{fig:ShannonTime}(a) and \ref{fig:ShannonTime}(b), we analyze the evolution of the Shannon entropy written in the basis corresponding to the eigenstates of the initial Hamiltonian,
\begin{equation}
\text{Sh}_{\text{ini}}^n (t) = - \sum_n W_n(t) \ln W_n(t) , 
\label{eq.sh1}
\end{equation}
where
\begin{equation}
W_n(t) =\left| \left\langle n \left| e^{-i \widehat{H}_{F} t} \right| \text{ini} \right\rangle \right|^2 \:
\end{equation}
is the probability to find the system in the basis vector $|n\rangle$ and $W_{\text{ini}}(t) = F(t)$.  
We focus on the quenches from $\widehat{H}_I$ to $\widehat{H}_F^{\text{local}}$ and from $\widehat{H}_I$ to $\widehat{H}_F^{\text{global}}$ in the intermediate perturbation regime.

The temporal behavior for both quenches is very similar. There is an initial quadratic growth, as
expected from perturbation theory, followed by a linear increase of the entropy, which is a general behavior of initial states that are sufficiently delocalized and can take place even when the final Hamiltonian is integrable~\cite{Flambaum2001b,Santos2012PRL,Santos2012PRE}.

A cascade model was developed to describe the progressive decay of the initial state into $|n\rangle$ \cite{Flambaum2001b}. It led to the following  semi-analytical expression for the Shannon entropy,
\begin{equation}
\text{Sh}_{\text{ini}}^n (t) = - F(t) \ln F(t) -
\left[ 1- F(t)\right] \ln \left( \frac{1- F(t)}{N_{pc}}\right),
\label{eq:Sanalytical}
\end{equation}
where $N_{pc} $ is the infinite time average of $\exp (\text{Sh}_{\text{ini}}^n)$. However, this expression [solid lines in Figs.~\ref{fig:ShannonTime}(a) and \ref{fig:ShannonTime}(b)] does not agree with our numerical expressions. The deviation happens because parity is still almost conserved in the initial integrable Hamiltonian. This is not the case for the two chaotic Hamiltonians and the reverse quenches, from the impurity and NNN Hamiltonians to the $XXZ$ system, lead to excellent agreement with Eq.~(\ref{eq:Sanalytical}) (not shown). 

The small defect on site 1 is not enough to strongly break reflection symmetry in the $XXZ$ model, so the initial states have significantly larger contributions from one of the parity sectors. This is deduced from Figs.~\ref{fig:ShannonTime}(c) and \ref{fig:ShannonTime}(d), where we separate the sum of the participations $W_n$ of quasi-even states from the sum of states with quasi-odd parity. The contributions from the states that do not belong to the same parity sector as $|\text{ini}\rangle$ take a very long time to become relevant. This results in a sort of pre-relaxation of the Shannon entropy. The fact that $N_{pc} $ is computed after relaxation causes the disagreement between Eq.~(\ref{eq:Sanalytical}) and the numerical results. Excellent agreement with the semi-analytical expression can be recovered, up to the pre-relaxation region, if $N_{pc} $ is computed by taking a time interval restricted to this region (not shown).

\section{Thermalization}
\label{Sec:Thermalization}

The subject of thermalization in isolated quantum systems~\cite{Jensen1985,Deutsch1991,Srednicki1994,Horoi1995,ZelevinskyRep1996,Flambaum1996b,Flambaum1997,Borgonovi1998,Borgonovi2000,I01} has been brought back to surface with recent experimental and theoretical studies~\cite{Kinoshita06,hofferberth07,Kollath2007,Manmana2007,Rigol2008,rigol09STATa,rigol09STATb}. A first necessary condition for thermalization is, of course, the relaxation of the quenched system to a new equilibrium.

One refers to equilibration in such isolated systems in a probabilistic sense. It occurs if after a transient time the system remains very close to a steady state  for most time and the fluctuations around it decreases with system size (see Ref.~\cite{Zangara2013} and references therein). In this scenario,  the density matrix,
\begin{eqnarray}
&&\rho(t) =
|\Psi(t)\rangle\langle\Psi(t)| 
\label{eq:rho}\\
&&= \!\sum_{\alpha} |C_{\alpha}^{\text{ini}}|^2 |\psi_{\alpha}\rangle \langle \psi_{\alpha}| \!+\! \sum_{\alpha \neq \beta} C_{\alpha}^{\text{ini*}} C_{\beta}^{\text{ini}} e^{i(E_\alpha-E_\beta)t}  |\psi_{\alpha}\rangle \langle \psi_{\beta}| ,
\nonumber
\end{eqnarray}
approaches the diagonal density matrix $\rho_{\rm DE}$ \cite{Polkovnikov2011,Santos2011PRL,Santos2012PRER}, which corresponds to the infinite time average,
\begin{equation}
\rho_{\rm DE} = \lim_{t\rightarrow \infty} \frac{1}{t}
\int^{t}_0 d\tau\,   \rho(\tau) =\sum_\alpha |C_{\alpha}^{\text{ini}}|^2 
|\psi_{\alpha}\rangle \langle \psi_{\alpha}|.
\label{eq:diagonal}
\end{equation}

The entropy that describes the system after relaxation,
\begin{equation}
S^{\alpha}_{\text{ini}}= -\text{Tr} (- \rho_{\rm DE} \ln \rho_{\rm DE} ) = - \sum_\alpha |C_{\alpha}^{\text{ini}}|^2 \ln |C_{\alpha}^{\text{ini}}|^2 ,
\label{eq:Sdiagonal}
\end{equation}
is referred to as diagonal entropy~\cite{Polkovnikov2011}. It is simply the Shannon entropy of the initial state projected onto the eigenstates of the final Hamiltonian. Thermalization occurs when the diagonal entropy coincides with the thermodynamic entropy~\cite{Polkovnikov2011,Santos2011PRL,Santos2012PRER}.

In terms of few-body observables, thermalization implies that their infinite time average (the diagonal ensemble average),
\begin{equation}
O_{\rm DE} = \text{Tr} ( \rho_{\rm DE}  \widehat{O} ) =\sum_\alpha |C_{\alpha}^{\text{ini}}|^2 \langle \psi_{\alpha} |\widehat{O} | \psi_{\alpha } \rangle ,
\label{eq:Ode}
\end{equation}
becomes very close to the thermal (microcanonical) average,
\begin{equation}
O_\text{ME}  \equiv  \frac{1}{{\cal{N}}_{E^{\text{ini}},\delta  E}}\hspace{-0.5cm}\sum_{\substack{\alpha \\ |E^{\text{ini}}-E_\alpha|<\delta E}}\hspace{-0.5cm} \langle \psi_{\alpha} |\widehat{O} | \psi_{\alpha } \rangle ,
\label{eq:micro}
\end{equation}
when the system is finite, and both averages coincide in the thermodynamic limit. Above ${\cal{N}}_{E^{\text{ini}},\delta E}$ stands for the number of energy eigenstates in the window $\delta E$.

Two situations can imply the proximity of the two averages:

(1) The eigenstate expectation value of the observables, $\langle \psi_{\alpha} |\widehat{O} | \psi_{\alpha } \rangle$, is a smooth function of energy, which means that the result from a single eigenstate inside the microcanonical window agrees with the microcanonical average. This approach became known as eigenstate thermalization hypothesis (ETH) \cite{Deutsch1991,Srednicki1994,Rigol2008,rigol09STATa,rigol09STATb}.

(2) The coefficients $C_{\alpha}^{\text{ini}}$ behave as random variables. This happens when the energy distribution of the initial state fills the energy shell. The fluctuations of the coefficients in Eq.~(\ref{eq:Ode}) become uncorrelated with $\langle \psi_{\alpha} |\widehat{O} | \psi_{\alpha } \rangle$ \cite{Flambaum1997}. 

For the quench into the impurity Hamiltonian, if thermalization happens, it must be mainly caused by condition (1), since the LDOS for this model does not achieve a Gaussian shape, as discussed in the end of Sec.~\ref{Sec:LDOS}.

\begin{figure}[htb]
\centering
\includegraphics*[width=0.48\textwidth]{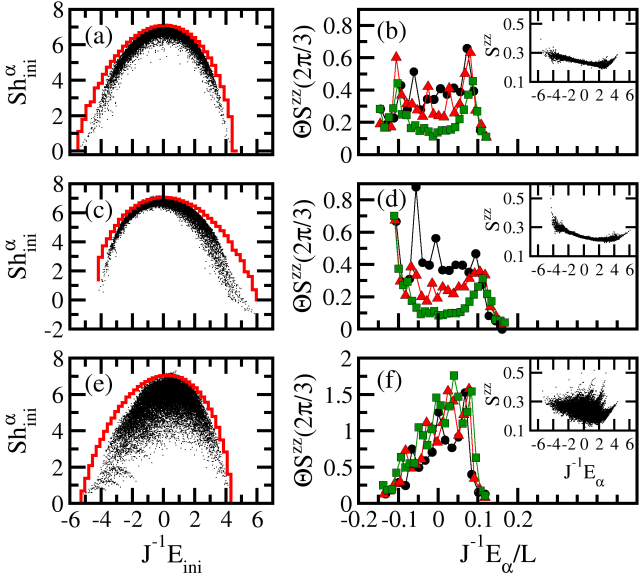}
\caption{(Color online) Left: Diagonal entropy (points) for all initial states and rescaled logarithm of the density of states for the final Hamiltonian (solid line). Right: Extremal fluctuations, $\Theta  S^{zz}$ [Eq.~(\ref{eq:MaxMin})] , for $\widehat{S}^{zz}(2\pi/3)$ in windows of energy $[E,E+0.4]$ (main panels) and the eigenstate expectation values of $\widehat{S}^{zz}(2\pi/3)$ for all $E_{\alpha}$ (insets). Final Hamiltonians: impurity (a), (b), NNN couplings (c), (d), $XXZ$ (e), (f). Initial Hamiltonians: $XXZ$ (a), (c) and $XX$ (e). $\Delta=0.48$, $d_F=0.9$, $\lambda_F=0.44$. Left panels and insets: $L=18$. Right main panels: $L=12$ (circles ); $L=15$ (triangles);  $L=18$ (squares).}
\label{fig:ETH}	
\end{figure}

\begin{figure}[htb]
\centering
\includegraphics*[width=0.48\textwidth]{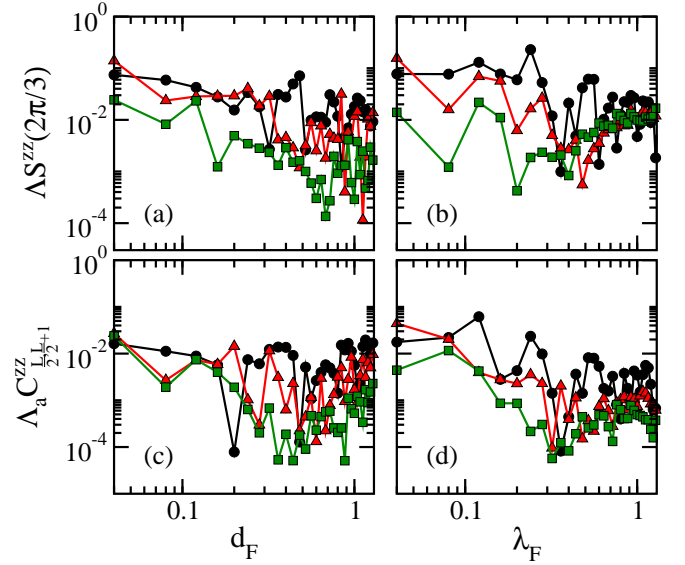}
\caption{(Color online) Relative difference for $\widehat{S}^{zz}(2\pi/3)$ (a), (b) and absolute difference for $C^{zz}_{L/2,L/2+1}$ (c), (d) vs $d_F$ (a), (c) and $\lambda_F$ (b), (d) Initial state with $T=7 J^{-1}$; $\Delta=0.48$; $L=12$ (circles ); $L=15$ (triangles);  $L=18$ (squares).}
\label{fig:SzzPerturbation}	
\end{figure}

\begin{figure}[htb]
\centering
\includegraphics*[width=0.48\textwidth]{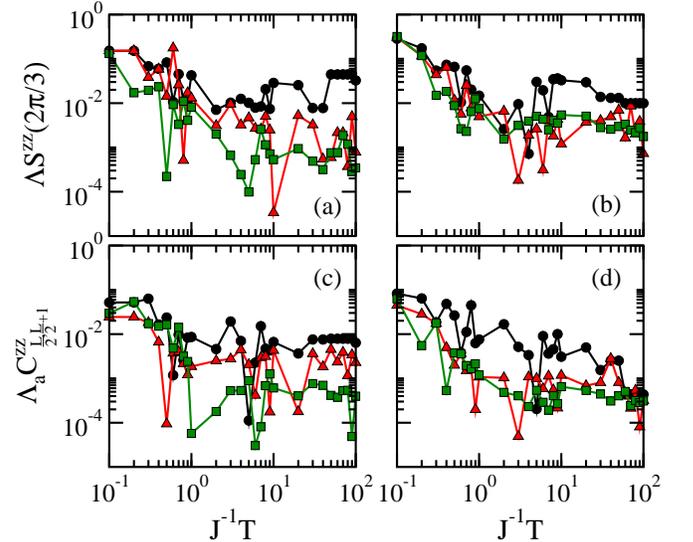}
\caption{(Color online) Relative difference for $\widehat{S}^{zz}(2\pi/3)$ (a), (b) and absolute difference for $C^{zz}_{L/2,L/2+1}$ (c), (d) vs temperature; $d_F=0.9$ (a), (c) and $\lambda_F=0.44$ (b), (d); $\Delta=0.48$; $L=12$ (circles ); $L=15$ (triangles);  $L=18$ (squares).}
\label{fig:SzzT}	
\end{figure}

In Fig.~\ref{fig:ETH}  we analyze the diagonal entropy and the eigenstates expectation values of few-body observables for the impurity, NNN, and $XXZ$ models. The results suggest agreement with ETH for the first two. In Figs.~\ref{fig:SzzPerturbation} and \ref{fig:SzzT} we confirm the expectations of thermalization for the local and global quenches into chaotic models.

Figures~\ref{fig:ETH}(a), \ref{fig:ETH}(c), and \ref{fig:ETH}(e) compare the diagonal entropy (points) for all the initial states extracted from $\widehat{H}_I$ with the logarithm of the density of states of the final Hamiltonian (solid line), which is related with the microcanonical entropy. Figures~\ref{fig:ETH}(a) and \ref{fig:ETH}(c) correspond to the quenches from the $XXZ$ model to the impurity and NNN Hamiltonians, respectively, and Fig.~\ref{fig:ETH}(e) is an integrable quench from the $XX$ model to the $XXZ$ Hamiltonian. In all panels, both entropies show a similar shape, apart from the edges of the spectrum. However, the fluctuations of the diagonal entropy in the integrable scenario is significantly larger and, contrary to the quenches into chaotic models, they do not decrease with system size~\cite{Santos2011PRL}.  This suggests that thermalization may not be possible when both Hamiltonians in the quench are integrable and the initial state has a finite temperature. This conjecture is confirmed by studies in the thermodynamic limit for quenches involving Hamiltonians that are of the $XXZ$-type before and after the perturbation~\cite{RigolARXIV}.

In Figs.~\ref{fig:ETH}(b), \ref{fig:ETH}(d), and \ref{fig:ETH}(f), we analyze the dependence on system size of the fluctuations of the eigenstate expectation values of the structure factor in the $z$ direction,
\begin{equation}
\widehat{S}^{zz}(k)=\frac{1}{L}\sum_{l,j=1}^L \widehat{S}_l^z \widehat{S}_j^z e^{-ik\left(l-j\right)} .
\label{eq:structure}
\end{equation}
This is a nonlocal observable in space. In the main panels we show the results of
\begin{equation}
\Theta  S^{zz} \equiv \left| \frac{\max S^{zz} - \min S^{zz}}{S^{zz}_\textrm{ME}} \right|
\label{eq:MaxMin}
\end{equation}
for three system sizes. Above, $\max S^{zz}$ ($\min S^{zz}$) stands for the maximum (minimum) value of $\langle \psi_{\alpha} |\widehat{S}^{zz}  | \psi_{\alpha } \rangle$ obtained in the energy window used to calculate $S^{zz}_\textrm{ME}$. This quantity measures the extremal fluctuations and is more appropriate to test ETH than the dispersion, which can decrease with $L$ simply because the number of states increases exponentially with $L$~\cite{Santos2010PREb}. The values of $\Theta  S^{zz}$ for the impurity and NNN models are comparable and, away from the edges of the spectrum, they clearly decreases with $L$. For the $XXZ$ system, $\Theta  S^{zz}$ is significantly larger and does not decrease with $L$. 
The insets in Fig.~\ref{fig:ETH} illustrate the fluctuations of the structure factor for a single system size. They show $\langle \psi_{\alpha} |\widehat{S}^{zz} | \psi_{\alpha } \rangle$ for all $E_{\alpha}$. The outcomes for the impurity and NNN Hamiltonians are equivalent and reflect the smooth behavior with energy of the Shannon entropy of the eigenstates $|\psi_{\alpha}\rangle$, seen in Fig.~\ref{fig:Ps} and typical of the chaotic domain \cite{Santos2010PRE,RigolSantos2010,Santos2010PREb}. This is in agreement with ETH and in contrast with the results for the $XXZ$ model, where large fluctuations are observed.

Figures~\ref{fig:SzzPerturbation} and \ref{fig:SzzT} compare the results for the infinite time averages and the microcanonical prediction. We compute the relative difference,
\begin{equation}
\Lambda S^{zz}=\frac{| S^{zz}_{\text{DE}}- S^{zz}_{\text{ME}}|}{|S^{zz}_{\text{DE}}|} ,
\label{eq:reldif}
\end{equation}
for the structure factor and also the absolute difference,
\begin{equation}
\Lambda_a C^{zz}=\left|C^{zz}_\text{DE}- C^{zz}_\text{ME}\right| ,
\end{equation}
for the spin-spin correlation in the $z$ direction for sites in the middle of the chain,
$\widehat{C}^{zz}_{L/2,L/2+1}= \langle \widehat{S}^z_{L/2} \widehat{S}^z_{L/2} \rangle$,
which is a local operator in space.

Figure~\ref{fig:SzzPerturbation} shows $\Lambda S^{zz}$ and $\Lambda_a C^{zz}$ for different values of the perturbations for the quenches to the impurity [Figs. \ref{fig:SzzPerturbation}(a) and \ref{fig:SzzPerturbation}(c)] and NNN [Figs. \ref{fig:SzzPerturbation}(b) and \ref{fig:SzzPerturbation}(d)] Hamiltonians. The relative differences are of similar magnitude for both models. Despite the large fluctuations, one can see that overall $\Lambda S^{zz}$ and $\Lambda_a C^{zz}$ decrease with the strength of the perturbation. This is better noticed by focusing on the largest system size, $L=18$ (green squares), where the fluctuations are smaller. The differences eventually start increasing again when $d_F$ and $\lambda_F$ get close to 1, since beyond this point another integrable limit is reached. By comparing most points for $L=12$ (black circles) with $L=18$, one also notices that $\Lambda S^{zz}$ and $\Lambda_a C^{zz}$ decrease with $L$.

Figure~\ref{fig:SzzT} shows $\Lambda S^{zz}$ and $\Lambda_a C^{zz}$ for different values of temperature for quenches to the impurity [Figs. \ref{fig:SzzT}(a) and \ref{fig:SzzT}(c)] and NNN [Figs. \ref{fig:SzzT}(b) and \ref{fig:SzzT}(d)] Hamiltonians. Despite the fluctuations, general trends can be identified. For both quenches, we see that the agreement between the averages improves as the temperature increases and  the energy of $|\text{ini}\rangle$ approaches the middle of the spectrum.  This is better seen if we concentrate our attention on $L=18$. The improvement with system size is also clear if we compare $L=12$  with $L=18$. The results reinforce the equivalence between the two models for the chosen values of $d_F$ and $\lambda_F$. They also corroborate the dependence on temperature in the studies of thermalization~\cite{He2013,Torres2013}. 

In Ref.~\cite{Torres2013}, we studied quenches to the NNN model in the limit of strong perturbation, where the energy distribution of the initial state is Gaussian and therefore already approximately thermal, even before the quench. In such an extreme scenario, it was difficult to clearly discern improvements with $T$ and $L$. Here, where the initial state is Breit-Wigner, these improvements are more visible.

\section{Summary}
\label{Sec:Summary}

We studied static and dynamical properties of one-dimensional spin-$\frac{1}{2}$ systems and showed that, for the initial states considered, the outcomes of a local-in-space quench can be very similar to those of a global quench. Our system was initially in an excited eigenstate of an integrable gapless $XXZ$ model. Either a local magnetic field was suddenly turned on or NNN couplings affecting the entire chain were instantaneously added.  Since the initial state is far from localized in space, it was globally affected by both perturbations. For both cases, we found that:

(i) The final Hamiltonian matrices have very similar structures if written in the mean-field basis corresponding to the eigenstates $|n\rangle$ of the initial $XXZ$ Hamiltonian. Their eigenvalues and eigenstates show equivalent statistical properties, which are typical of quantum systems in the chaotic domain. 

(ii) The energy distribution of initial states away from the borders of the spectrum have a Breit-Wigner form and lead to an exponential decay of the fidelity. 

(iii) The Shannon entropy written in the basis $|n\rangle $ grows linearly in time and reaches a prethermalization region caused by parity (almost) conservation, before saturating.

(iv) In agreement with ETH, the eigenstate expectation values of few-body observables show minor fluctuations for states close in energy and, away from the edges of the spectrum, they decrease with system size. The viability of thermalization is confirmed by comparing infinite time averages and microcanonical predictions. The results are similar and further improve with $L$.  

A main goal of this paper was to show that the effects of a quench depend on the representation and therefore on the initial state. A perturbation that is local in a certain basis should have limited effects and cause a restricted propagation of the signal if the initial state is also localized in that same basis, as shown for example in \cite{Fukuhara2013,Calabrese2007}. But in our studies, the initial states considered were delocalized in space, so it is not so surprising that the local quench in space turned out to be very similar to the global one.

There is, however, an important difference between the two quenches analyzed. As $\lambda_F$ increases, the global quench can eventually reach the strong perturbation regime, where both the energy distribution of the initial state and the fidelity decay become Gaussian. This extreme is not achieved by the local quench of the magnetic field (at least not for the system sizes accessible to exact diagonalization). It is conceivable that such local quenches be fundamentally forbidden to reach this regime. This is a point that deserves further investigation.

\begin{acknowledgments}
This work was supported by the  NSF grant No.~DMR-1147430. E.J.T.H. acknowledges partial support from CONACyT, Mexico. L. F. S. thanks Tony Apollaro for discussions that inspired this work.
\end{acknowledgments}


\begin{thebibliography}{88}
\expandafter\ifx\csname natexlab\endcsname\relax\def\natexlab#1{#1}\fi
\expandafter\ifx\csname bibnamefont\endcsname\relax
  \def\bibnamefont#1{#1}\fi
\expandafter\ifx\csname bibfnamefont\endcsname\relax
  \def\bibfnamefont#1{#1}\fi
\expandafter\ifx\csname citenamefont\endcsname\relax
  \def\citenamefont#1{#1}\fi
\expandafter\ifx\csname url\endcsname\relax
  \def\url#1{\texttt{#1}}\fi
\expandafter\ifx\csname urlprefix\endcsname\relax\def\urlprefix{URL }\fi
\providecommand{\bibinfo}[2]{#2}
\providecommand{\eprint}[2][]{\url{#2}}

\bibitem[{\citenamefont{Cazalilla et~al.}(2011)\citenamefont{Cazalilla, Citro,
  Giamarchi, Orignac, and Rigol}}]{Cazalilla2011}
\bibinfo{author}{\bibfnamefont{M.~A.} \bibnamefont{Cazalilla}},
  \bibinfo{author}{\bibfnamefont{R.}~\bibnamefont{Citro}},
  \bibinfo{author}{\bibfnamefont{T.}~\bibnamefont{Giamarchi}},
  \bibinfo{author}{\bibfnamefont{E.}~\bibnamefont{Orignac}}, \bibnamefont{and}
  \bibinfo{author}{\bibfnamefont{M.}~\bibnamefont{Rigol}},
  \bibinfo{journal}{Rev. Mod. Phys.} \textbf{\bibinfo{volume}{83}},
  \bibinfo{pages}{1405 } (\bibinfo{year}{2011}).

\bibitem[{\citenamefont{Polkovnikov et~al.}(2011)\citenamefont{Polkovnikov,
  Sengupta, Silva, and Vengalattore}}]{Polkovnikov2011RMP}
\bibinfo{author}{\bibfnamefont{A.}~\bibnamefont{Polkovnikov}},
  \bibinfo{author}{\bibfnamefont{K.}~\bibnamefont{Sengupta}},
  \bibinfo{author}{\bibfnamefont{A.}~\bibnamefont{Silva}}, \bibnamefont{and}
  \bibinfo{author}{\bibfnamefont{M.}~\bibnamefont{Vengalattore}},
  \bibinfo{journal}{Rev. Mod. Phys.} \textbf{\bibinfo{volume}{83}},
  \bibinfo{pages}{863} (\bibinfo{year}{2011}).

\bibitem[{\citenamefont{White and Feiguin}(2004)}]{White2004}
\bibinfo{author}{\bibfnamefont{S.~R.} \bibnamefont{White}} \bibnamefont{and}
  \bibinfo{author}{\bibfnamefont{A.~E.} \bibnamefont{Feiguin}},
  \bibinfo{journal}{Phys. Rev. Lett.} \textbf{\bibinfo{volume}{93}},
  \bibinfo{pages}{076401} (\bibinfo{year}{2004}).

\bibitem[{\citenamefont{Schollw{\"o}ck}(2013)}]{Schollwock2013}
\bibinfo{author}{\bibfnamefont{U.}~\bibnamefont{Schollw{\"o}ck}}, in
  \emph{\bibinfo{booktitle}{Strongly Correlated Systems: Numerical Methods}},
  edited by \bibinfo{editor}{\bibfnamefont{A.}~\bibnamefont{Avella}}
  \bibnamefont{and} \bibinfo{editor}{\bibfnamefont{F.}~\bibnamefont{Mancini}}
  (\bibinfo{publisher}{Springer Verlag}, \bibinfo{address}{Berlin Heidelberg},
  \bibinfo{year}{2013}).

\bibitem[{\citenamefont{Rigol et~al.}(2006)\citenamefont{Rigol, Bryant, and
  Singh}}]{Rigol2006}
\bibinfo{author}{\bibfnamefont{M.}~\bibnamefont{Rigol}},
  \bibinfo{author}{\bibfnamefont{T.}~\bibnamefont{Bryant}}, \bibnamefont{and}
  \bibinfo{author}{\bibfnamefont{R.~R.~P.} \bibnamefont{Singh}},
  \bibinfo{journal}{Phys. Rev. Lett.} \textbf{\bibinfo{volume}{97}},
  \bibinfo{pages}{187202} (\bibinfo{year}{2006}).

\bibitem[{\citenamefont{Rigol}()}]{RigolARXIV}
\bibinfo{author}{\bibfnamefont{M.}~\bibnamefont{Rigol}},
  \bibinfo{note}{arXiv:1401.2160}.

\bibitem[{\citenamefont{Cappellaro et~al.}(2007)\citenamefont{Cappellaro,
  Ramanathan, and Cory}}]{Cappellaro2007}
\bibinfo{author}{\bibfnamefont{P.}~\bibnamefont{Cappellaro}},
  \bibinfo{author}{\bibfnamefont{C.}~\bibnamefont{Ramanathan}},
  \bibnamefont{and} \bibinfo{author}{\bibfnamefont{D.~G.} \bibnamefont{Cory}},
  \bibinfo{journal}{Phys. Rev. Lett.} \textbf{\bibinfo{volume}{99}},
  \bibinfo{pages}{250506 (1} (\bibinfo{year}{2007}).

\bibitem[{\citenamefont{Ramanathan et~al.}(2011)\citenamefont{Ramanathan,
  Cappellaro, Viola, and Cory}}]{Ramanathan2011}
\bibinfo{author}{\bibfnamefont{C.}~\bibnamefont{Ramanathan}},
  \bibinfo{author}{\bibfnamefont{P.}~\bibnamefont{Cappellaro}},
  \bibinfo{author}{\bibfnamefont{L.}~\bibnamefont{Viola}}, \bibnamefont{and}
  \bibinfo{author}{\bibfnamefont{D.}~\bibnamefont{Cory}}, \bibinfo{journal}{New
  J. Phys.} \textbf{\bibinfo{volume}{13}}, \bibinfo{pages}{103015}
  (\bibinfo{year}{2011}).

\bibitem[{\citenamefont{Kaur et~al.}()\citenamefont{Kaur, Ajoy, and
  Cappellaro}}]{KaurARXIV}
\bibinfo{author}{\bibfnamefont{G.}~\bibnamefont{Kaur}},
  \bibinfo{author}{\bibfnamefont{A.}~\bibnamefont{Ajoy}}, \bibnamefont{and}
  \bibinfo{author}{\bibfnamefont{P.}~\bibnamefont{Cappellaro}},
\bibinfo{journal}{New
  J. Phys.} \textbf{\bibinfo{volume}{15}}, \bibinfo{pages}{093035}
  (\bibinfo{year}{2013}).

\bibitem[{\citenamefont{Greiner et~al.}(2002)\citenamefont{Greiner, Mandel,
  Esslinger, H\"ansch, and Bloch}}]{Greiner2002}
\bibinfo{author}{\bibfnamefont{M.}~\bibnamefont{Greiner}},
  \bibinfo{author}{\bibfnamefont{O.}~\bibnamefont{Mandel}},
  \bibinfo{author}{\bibfnamefont{T.}~\bibnamefont{Esslinger}},
  \bibinfo{author}{\bibfnamefont{T.~W.} \bibnamefont{H\"ansch}},
  \bibnamefont{and} \bibinfo{author}{\bibfnamefont{I.}~\bibnamefont{Bloch}},
  \bibinfo{journal}{Nature} \textbf{\bibinfo{volume}{415}}, \bibinfo{pages}{39}
  (\bibinfo{year}{2002}).

\bibitem[{\citenamefont{Kinoshita et~al.}(2006)\citenamefont{Kinoshita, Wenger,
  and Weiss}}]{Kinoshita06}
\bibinfo{author}{\bibfnamefont{T.}~\bibnamefont{Kinoshita}},
  \bibinfo{author}{\bibfnamefont{T.}~\bibnamefont{Wenger}}, \bibnamefont{and}
  \bibinfo{author}{\bibfnamefont{D.~S.} \bibnamefont{Weiss}},
  \bibinfo{journal}{Nature} \textbf{\bibinfo{volume}{440}},
  \bibinfo{pages}{900} (\bibinfo{year}{2006}).

\bibitem[{\citenamefont{Hofferberth et~al.}(2007)\citenamefont{Hofferberth,
  Lesanovsky, Fischer, Schumm, and Schmiedmayer}}]{hofferberth07}
\bibinfo{author}{\bibfnamefont{S.}~\bibnamefont{Hofferberth}},
  \bibinfo{author}{\bibfnamefont{I.}~\bibnamefont{Lesanovsky}},
  \bibinfo{author}{\bibfnamefont{B.}~\bibnamefont{Fischer}},
  \bibinfo{author}{\bibfnamefont{T.}~\bibnamefont{Schumm}}, \bibnamefont{and}
  \bibinfo{author}{\bibfnamefont{J.}~\bibnamefont{Schmiedmayer}},
  \bibinfo{journal}{Nature} \textbf{\bibinfo{volume}{449}},
  \bibinfo{pages}{324} (\bibinfo{year}{2007}).

\bibitem[{\citenamefont{Bloch et~al.}(2008)\citenamefont{Bloch, Dalibard, and
  Zwerger}}]{Bloch2008}
\bibinfo{author}{\bibfnamefont{I.}~\bibnamefont{Bloch}},
  \bibinfo{author}{\bibfnamefont{J.}~\bibnamefont{Dalibard}}, \bibnamefont{and}
  \bibinfo{author}{\bibfnamefont{W.}~\bibnamefont{Zwerger}},
  \bibinfo{journal}{Rev. Mod. Phys.} \textbf{\bibinfo{volume}{80}},
  \bibinfo{pages}{885} (\bibinfo{year}{2008}).

\bibitem[{\citenamefont{Trotzky et~al.}(2008)\citenamefont{Trotzky, Cheinet,
  F\"olling, Feld, Schnorrberger, Rey, Polkovnikov, Demler, Lukin, and
  Bloch}}]{Trotzky2008}
\bibinfo{author}{\bibfnamefont{S.}~\bibnamefont{Trotzky}},
  \bibinfo{author}{\bibfnamefont{P.}~\bibnamefont{Cheinet}},
  \bibinfo{author}{\bibfnamefont{S.}~\bibnamefont{F\"olling}},
  \bibinfo{author}{\bibfnamefont{M.}~\bibnamefont{Feld}},
  \bibinfo{author}{\bibfnamefont{U.}~\bibnamefont{Schnorrberger}},
  \bibinfo{author}{\bibfnamefont{A.~M.} \bibnamefont{Rey}},
  \bibinfo{author}{\bibfnamefont{A.}~\bibnamefont{Polkovnikov}},
  \bibinfo{author}{\bibfnamefont{E.~A.} \bibnamefont{Demler}},
  \bibinfo{author}{\bibfnamefont{M.~D.} \bibnamefont{Lukin}}, \bibnamefont{and}
  \bibinfo{author}{\bibfnamefont{I.}~\bibnamefont{Bloch}},
  \bibinfo{journal}{Science} \textbf{\bibinfo{volume}{319}},
  \bibinfo{pages}{295} (\bibinfo{year}{2008}).

\bibitem[{\citenamefont{Simon et~al.}(2011)\citenamefont{Simon, Bakr, Ma, Tai,
  Preiss, and Greiner}}]{Simon2011}
\bibinfo{author}{\bibfnamefont{J.}~\bibnamefont{Simon}},
  \bibinfo{author}{\bibfnamefont{W.~S.} \bibnamefont{Bakr}},
  \bibinfo{author}{\bibfnamefont{R.}~\bibnamefont{Ma}},
  \bibinfo{author}{\bibfnamefont{M.~E.} \bibnamefont{Tai}},
  \bibinfo{author}{\bibfnamefont{P.~M.} \bibnamefont{Preiss}},
  \bibnamefont{and} \bibinfo{author}{\bibfnamefont{M.}~\bibnamefont{Greiner}},
  \bibinfo{journal}{Nature (London)} \textbf{\bibinfo{volume}{472}},
  \bibinfo{pages}{307} (\bibinfo{year}{2011}).

\bibitem[{\citenamefont{Trotzky et~al.}(2012)\citenamefont{Trotzky, Chen,
  Flesch, McCulloch, Schollw\"ock, Eisert, and Bloch}}]{Trotzky2012}
\bibinfo{author}{\bibfnamefont{S.}~\bibnamefont{Trotzky}},
  \bibinfo{author}{\bibfnamefont{Y.-A.} \bibnamefont{Chen}},
  \bibinfo{author}{\bibfnamefont{A.}~\bibnamefont{Flesch}},
  \bibinfo{author}{\bibfnamefont{I.~P.} \bibnamefont{McCulloch}},
  \bibinfo{author}{\bibfnamefont{U.}~\bibnamefont{Schollw\"ock}},
  \bibinfo{author}{\bibfnamefont{J.}~\bibnamefont{Eisert}}, \bibnamefont{and}
  \bibinfo{author}{\bibfnamefont{I.}~\bibnamefont{Bloch}},
  \bibinfo{journal}{Nature Phys.} \textbf{\bibinfo{volume}{8}},
  \bibinfo{pages}{325} (\bibinfo{year}{2012}).

\bibitem[{\citenamefont{Fukuhara et~al.}(2013)\citenamefont{Fukuhara, Kantian,
  Endres, Cheneau, Schausz, Hild, Bellem, Schollw\"ock, Giamarchi, Gross
  et~al.}}]{Fukuhara2013}
\bibinfo{author}{\bibfnamefont{T.}~\bibnamefont{Fukuhara}},
  \bibinfo{author}{\bibfnamefont{A.}~\bibnamefont{Kantian}},
  \bibinfo{author}{\bibfnamefont{M.}~\bibnamefont{Endres}},
  \bibinfo{author}{\bibfnamefont{M.}~\bibnamefont{Cheneau}},
  \bibinfo{author}{\bibfnamefont{P.}~\bibnamefont{Schausz}},
  \bibinfo{author}{\bibfnamefont{S.}~\bibnamefont{Hild}},
  \bibinfo{author}{\bibfnamefont{D.}~\bibnamefont{Bellem}},
  \bibinfo{author}{\bibfnamefont{U.}~\bibnamefont{Schollw\"ock}},
  \bibinfo{author}{\bibfnamefont{T.}~\bibnamefont{Giamarchi}},
  \bibinfo{author}{\bibfnamefont{C.}~\bibnamefont{Gross}},
  \bibnamefont{et~al.}, \bibinfo{journal}{Nat. Phys.}
  \textbf{\bibinfo{volume}{9}}, \bibinfo{pages}{235} (\bibinfo{year}{2013}).

\bibitem[{\citenamefont{Calabrese and Cardy}(2007)}]{Calabrese2007}
\bibinfo{author}{\bibfnamefont{P.}~\bibnamefont{Calabrese}} \bibnamefont{and}
  \bibinfo{author}{\bibfnamefont{J.}~\bibnamefont{Cardy}}, \bibinfo{journal}{J.
  Stat. Mech.} \textbf{\bibinfo{volume}{2007}}, \bibinfo{pages}{P10004}
  (\bibinfo{year}{2007}).

\bibitem[{\citenamefont{Eisler and Peschel}(2007)}]{Eisler2007}
\bibinfo{author}{\bibfnamefont{V.}~\bibnamefont{Eisler}} \bibnamefont{and}
  \bibinfo{author}{\bibfnamefont{I.}~\bibnamefont{Peschel}},
  \bibinfo{journal}{J. Stat. Mech.} \textbf{\bibinfo{volume}{2007}},
  \bibinfo{pages}{P06005} (\bibinfo{year}{2007}).

\bibitem[{\citenamefont{Eisler et~al.}(2008)\citenamefont{Eisler, Karevski,
  Platini, and Peschel}}]{Eisler2008}
\bibinfo{author}{\bibfnamefont{V.}~\bibnamefont{Eisler}},
  \bibinfo{author}{\bibfnamefont{D.}~\bibnamefont{Karevski}},
  \bibinfo{author}{\bibfnamefont{T.}~\bibnamefont{Platini}}, \bibnamefont{and}
  \bibinfo{author}{\bibfnamefont{I.}~\bibnamefont{Peschel}},
  \bibinfo{journal}{J. Stat. Mech.} \textbf{\bibinfo{volume}{2008}},
  \bibinfo{pages}{P01023} (\bibinfo{year}{2008}).

\bibitem[{\citenamefont{Calabrese and Cardy}(2009)}]{Calabrese2009}
\bibinfo{author}{\bibfnamefont{P.}~\bibnamefont{Calabrese}} \bibnamefont{and}
  \bibinfo{author}{\bibfnamefont{J.}~\bibnamefont{Cardy}}, \bibinfo{journal}{J.
  Math. Phys. A} \textbf{\bibinfo{volume}{42}}, \bibinfo{pages}{4005}
  (\bibinfo{year}{2009}).

\bibitem[{\citenamefont{Diez et~al.}(2010)\citenamefont{Diez, Chancellor, Haas,
  Venuti, and Zanardi}}]{Diez2010}
\bibinfo{author}{\bibfnamefont{M.}~\bibnamefont{Diez}},
  \bibinfo{author}{\bibfnamefont{N.}~\bibnamefont{Chancellor}},
  \bibinfo{author}{\bibfnamefont{S.}~\bibnamefont{Haas}},
  \bibinfo{author}{\bibfnamefont{L.~C.} \bibnamefont{Venuti}},
  \bibnamefont{and} \bibinfo{author}{\bibfnamefont{P.}~\bibnamefont{Zanardi}},
  \bibinfo{journal}{Phys. Rev. A} \textbf{\bibinfo{volume}{82}},
  \bibinfo{pages}{032113} (\bibinfo{year}{2010}).

\bibitem[{\citenamefont{St\'ephan and Dubail}(2011)}]{Stephan2011}
\bibinfo{author}{\bibfnamefont{J.-M.} \bibnamefont{St\'ephan}}
  \bibnamefont{and} \bibinfo{author}{\bibfnamefont{J.}~\bibnamefont{Dubail}},
  \bibinfo{journal}{J. Stat. Mech.} \textbf{\bibinfo{volume}{2011}},
  \bibinfo{pages}{P08019} (\bibinfo{year}{2011}).

\bibitem[{\citenamefont{Divakaran et~al.}(2011)\citenamefont{Divakaran,
  Igl\'oi, and Rieger}}]{Divakaran2011}
\bibinfo{author}{\bibfnamefont{U.}~\bibnamefont{Divakaran}},
  \bibinfo{author}{\bibfnamefont{F.}~\bibnamefont{Igl\'oi}}, \bibnamefont{and}
  \bibinfo{author}{\bibfnamefont{H.}~\bibnamefont{Rieger}},
  \bibinfo{journal}{J. Stat. Mech.} \textbf{\bibinfo{volume}{2011}},
  \bibinfo{pages}{P10027} (\bibinfo{year}{2011}).

\bibitem[{\citenamefont{Smacchia and Silva}(2012)}]{Smacchia2012}
\bibinfo{author}{\bibfnamefont{P.}~\bibnamefont{Smacchia}} \bibnamefont{and}
  \bibinfo{author}{\bibfnamefont{A.}~\bibnamefont{Silva}},
  \bibinfo{journal}{Phys. Rev. Lett.} \textbf{\bibinfo{volume}{109}},
  \bibinfo{pages}{037202} (\bibinfo{year}{2012}).

\bibitem[{\citenamefont{Nozaki et~al.}(2013)\citenamefont{Nozaki, Numasawa, and
  Takayanagi}}]{Nozaki2013}
\bibinfo{author}{\bibfnamefont{M.}~\bibnamefont{Nozaki}},
  \bibinfo{author}{\bibfnamefont{T.}~\bibnamefont{Numasawa}}, \bibnamefont{and}
  \bibinfo{author}{\bibfnamefont{T.}~\bibnamefont{Takayanagi}},
  \bibinfo{journal}{J. High Energ. Phys.} \textbf{\bibinfo{volume}{2013}},
  \bibinfo{pages}{1} (\bibinfo{year}{2013}).

\bibitem[{\citenamefont{Asplund and Bernamonti}()}]{AsplundARXIV}
\bibinfo{author}{\bibfnamefont{C.~T.} \bibnamefont{Asplund}} \bibnamefont{and}
  \bibinfo{author}{\bibfnamefont{A.}~\bibnamefont{Bernamonti}},
  \bibinfo{note}{arXiv:1311.4173}.

\bibitem[{\citenamefont{Alba and Heidrich-Meisner}()}]{AlbaARXIV}
\bibinfo{author}{\bibfnamefont{V.}~\bibnamefont{Alba}} \bibnamefont{and}
  \bibinfo{author}{\bibfnamefont{F.}~\bibnamefont{Heidrich-Meisner}},
\bibinfo{journal}{Phys. Rev. B}
  \textbf{\bibinfo{volume}{90}}, \bibinfo{pages}{075144}
  (\bibinfo{year}{2014}).

\bibitem[{\citenamefont{Ganahl et~al.}(2012)\citenamefont{Ganahl, Rabel,
  Essler, and Evertz}}]{Ganahl2012}
\bibinfo{author}{\bibfnamefont{M.}~\bibnamefont{Ganahl}},
  \bibinfo{author}{\bibfnamefont{E.}~\bibnamefont{Rabel}},
  \bibinfo{author}{\bibfnamefont{F.~H.~L.} \bibnamefont{Essler}},
  \bibnamefont{and} \bibinfo{author}{\bibfnamefont{H.~G.}
  \bibnamefont{Evertz}}, \bibinfo{journal}{Phys. Rev. Lett.}
  \textbf{\bibinfo{volume}{108}}, \bibinfo{pages}{077206}
  (\bibinfo{year}{2012}).

\bibitem[{\citenamefont{Santos}(2004)}]{Santos2004}
\bibinfo{author}{\bibfnamefont{L.~F.} \bibnamefont{Santos}},
  \bibinfo{journal}{J. Phys. A} \textbf{\bibinfo{volume}{37}},
  \bibinfo{pages}{4723} (\bibinfo{year}{2004}).

\bibitem[{\citenamefont{Barisic et~al.}(2009)\citenamefont{Barisic,
  Prelov\v{s}ek, Metavitsiadis, and Zotos}}]{Barisic2009}
\bibinfo{author}{\bibfnamefont{O.~S.} \bibnamefont{Barisic}},
  \bibinfo{author}{\bibfnamefont{P.}~\bibnamefont{Prelov\v{s}ek}},
  \bibinfo{author}{\bibfnamefont{A.}~\bibnamefont{Metavitsiadis}},
  \bibnamefont{and} \bibinfo{author}{\bibfnamefont{X.}~\bibnamefont{Zotos}},
  \bibinfo{journal}{Phys. Rev. B} \textbf{\bibinfo{volume}{80}},
  \bibinfo{pages}{125118} (\bibinfo{year}{2009}).

\bibitem[{\citenamefont{Santos and Mitra}(2011)}]{Santos2011}
\bibinfo{author}{\bibfnamefont{L.~F.} \bibnamefont{Santos}} \bibnamefont{and}
  \bibinfo{author}{\bibfnamefont{A.}~\bibnamefont{Mitra}},
  \bibinfo{journal}{Phys. Rev. E} \textbf{\bibinfo{volume}{84}},
  \bibinfo{pages}{016206} (\bibinfo{year}{2011}).

\bibitem[{\citenamefont{Sinai}(1970)}]{Sinai1970}
\bibinfo{author}{\bibfnamefont{Y.~G.} \bibnamefont{Sinai}},
  \bibinfo{journal}{Russ. Math. Surv.} \textbf{\bibinfo{volume}{25}},
  \bibinfo{pages}{137} (\bibinfo{year}{1970}).

\bibitem[{\citenamefont{Berry}(1981)}]{Berry1981}
\bibinfo{author}{\bibfnamefont{M.~V.} \bibnamefont{Berry}},
  \bibinfo{journal}{Ann. Phys. (NY)} \textbf{\bibinfo{volume}{131}},
  \bibinfo{pages}{163} (\bibinfo{year}{1981}).

\bibitem[{\citenamefont{Zelevinsky et~al.}(1996)\citenamefont{Zelevinsky,
  Brown, Frazier, and Horoi}}]{ZelevinskyRep1996}
\bibinfo{author}{\bibfnamefont{V.}~\bibnamefont{Zelevinsky}},
  \bibinfo{author}{\bibfnamefont{B.~A.} \bibnamefont{Brown}},
  \bibinfo{author}{\bibfnamefont{N.}~\bibnamefont{Frazier}}, \bibnamefont{and}
  \bibinfo{author}{\bibfnamefont{M.}~\bibnamefont{Horoi}},
  \bibinfo{journal}{Phys. Rep.} \textbf{\bibinfo{volume}{276}},
  \bibinfo{pages}{85} (\bibinfo{year}{1996}).

\bibitem[{\citenamefont{Jacquod et~al.}(2001)\citenamefont{Jacquod, Silvestrov,
  and Beenakker}}]{Jacquod2001}
\bibinfo{author}{\bibfnamefont{P.}~\bibnamefont{Jacquod}},
  \bibinfo{author}{\bibfnamefont{P.}~\bibnamefont{Silvestrov}},
  \bibnamefont{and}
  \bibinfo{author}{\bibfnamefont{C.}~\bibnamefont{Beenakker}},
  \bibinfo{journal}{Phys. Rev. E} \textbf{\bibinfo{volume}{64}},
  \bibinfo{pages}{055203} (\bibinfo{year}{2001}).

\bibitem[{\citenamefont{Flambaum and
  Izrailev}(2001{\natexlab{a}})}]{Flambaum2001a}
\bibinfo{author}{\bibfnamefont{V.~V.} \bibnamefont{Flambaum}} \bibnamefont{and}
  \bibinfo{author}{\bibfnamefont{F.~M.} \bibnamefont{Izrailev}},
  \bibinfo{journal}{Phys. Rev. E} \textbf{\bibinfo{volume}{64}},
  \bibinfo{pages}{026124} (\bibinfo{year}{2001}{\natexlab{a}}).

\bibitem[{\citenamefont{Flambaum and
  Izrailev}(2001{\natexlab{b}})}]{Flambaum2001b}
\bibinfo{author}{\bibfnamefont{V.~V.} \bibnamefont{Flambaum}} \bibnamefont{and}
  \bibinfo{author}{\bibfnamefont{F.~M.} \bibnamefont{Izrailev}},
  \bibinfo{journal}{Phys. Rev. E} \textbf{\bibinfo{volume}{64}},
  \bibinfo{pages}{036220} (\bibinfo{year}{2001}{\natexlab{b}}).

\bibitem[{\citenamefont{Emerson et~al.}(2002)\citenamefont{Emerson, Weinstein,
  Lloyd, and Cory}}]{Emerson2002}
\bibinfo{author}{\bibfnamefont{J.}~\bibnamefont{Emerson}},
  \bibinfo{author}{\bibfnamefont{Y.~S.} \bibnamefont{Weinstein}},
  \bibinfo{author}{\bibfnamefont{S.}~\bibnamefont{Lloyd}}, \bibnamefont{and}
  \bibinfo{author}{\bibfnamefont{D.~G.} \bibnamefont{Cory}},
  \bibinfo{journal}{Phys. Rev. Lett.} \textbf{\bibinfo{volume}{89}},
  \bibinfo{pages}{284102} (\bibinfo{year}{2002}).

\bibitem[{\citenamefont{Weinstein et~al.}(2003)\citenamefont{Weinstein,
  Emerson, Lloyd, and Cory}}]{Weinstein2003}
\bibinfo{author}{\bibfnamefont{Y.~S.} \bibnamefont{Weinstein}},
  \bibinfo{author}{\bibfnamefont{J.}~\bibnamefont{Emerson}},
  \bibinfo{author}{\bibfnamefont{S.}~\bibnamefont{Lloyd}}, \bibnamefont{and}
  \bibinfo{author}{\bibfnamefont{D.}~\bibnamefont{Cory}},
  \bibinfo{journal}{Quant. Inf. Proc.} \textbf{\bibinfo{volume}{1}},
  \bibinfo{pages}{439} (\bibinfo{year}{2003}).

\bibitem[{\citenamefont{Izrailev and
  Casta{\~{n}}eda-Mendoza}(2006)}]{Izrailev2006}
\bibinfo{author}{\bibfnamefont{F.~M.} \bibnamefont{Izrailev}} \bibnamefont{and}
  \bibinfo{author}{\bibfnamefont{A.}~\bibnamefont{Casta{\~{n}}eda-Mendoza}},
  \bibinfo{journal}{Phys. Lett. A} \textbf{\bibinfo{volume}{350}},
  \bibinfo{pages}{355} (\bibinfo{year}{2006}).

\bibitem[{\citenamefont{Santos et~al.}(2012{\natexlab{a}})\citenamefont{Santos,
  Borgonovi, and Izrailev}}]{Santos2012PRL}
\bibinfo{author}{\bibfnamefont{L.~F.} \bibnamefont{Santos}},
  \bibinfo{author}{\bibfnamefont{F.}~\bibnamefont{Borgonovi}},
  \bibnamefont{and} \bibinfo{author}{\bibfnamefont{F.~M.}
  \bibnamefont{Izrailev}}, \bibinfo{journal}{Phys. Rev. Lett.}
  \textbf{\bibinfo{volume}{108}}, \bibinfo{pages}{094102}
  (\bibinfo{year}{2012}{\natexlab{a}}).

\bibitem[{\citenamefont{Santos et~al.}(2012{\natexlab{b}})\citenamefont{Santos,
  Borgonovi, and Izrailev}}]{Santos2012PRE}
\bibinfo{author}{\bibfnamefont{L.~F.} \bibnamefont{Santos}},
  \bibinfo{author}{\bibfnamefont{F.}~\bibnamefont{Borgonovi}},
  \bibnamefont{and} \bibinfo{author}{\bibfnamefont{F.~M.}
  \bibnamefont{Izrailev}}, \bibinfo{journal}{Phys. Rev. E}
  \textbf{\bibinfo{volume}{85}}, \bibinfo{pages}{036209}
  (\bibinfo{year}{2012}{\natexlab{b}}).

\bibitem[{\citenamefont{Flambaum et~al.}(1994)\citenamefont{Flambaum,
  Gribakina, Gribakin, and Kozlov}}]{Flambaum1994}
\bibinfo{author}{\bibfnamefont{V.~V.} \bibnamefont{Flambaum}},
  \bibinfo{author}{\bibfnamefont{A.~A.} \bibnamefont{Gribakina}},
  \bibinfo{author}{\bibfnamefont{G.~F.} \bibnamefont{Gribakin}},
  \bibnamefont{and} \bibinfo{author}{\bibfnamefont{M.~G.}
  \bibnamefont{Kozlov}}, \bibinfo{journal}{Phys. Rev. A}
  \textbf{\bibinfo{volume}{50}}, \bibinfo{pages}{267} (\bibinfo{year}{1994}).

\bibitem[{\citenamefont{Frazier et~al.}(1996)\citenamefont{Frazier, Brown, and
  Zelevinsky}}]{Frazier1996}
\bibinfo{author}{\bibfnamefont{N.}~\bibnamefont{Frazier}},
  \bibinfo{author}{\bibfnamefont{B.~A.} \bibnamefont{Brown}}, \bibnamefont{and}
  \bibinfo{author}{\bibfnamefont{V.}~\bibnamefont{Zelevinsky}},
  \bibinfo{journal}{Phys. Rev. C} \textbf{\bibinfo{volume}{54}},
  \bibinfo{pages}{1665} (\bibinfo{year}{1996}).

\bibitem[{\citenamefont{Flambaum and Izrailev}(2000)}]{Flambaum2000}
\bibinfo{author}{\bibfnamefont{V.~V.} \bibnamefont{Flambaum}} \bibnamefont{and}
  \bibinfo{author}{\bibfnamefont{F.~M.} \bibnamefont{Izrailev}},
  \bibinfo{journal}{Phys. Rev. E} \textbf{\bibinfo{volume}{61}},
  \bibinfo{pages}{2539} (\bibinfo{year}{2000}).

\bibitem[{\citenamefont{Torres-Herrera and Santos}()}]{Torres2014PRA}
\bibinfo{author}{\bibfnamefont{E.~J.} \bibnamefont{Torres-Herrera}}
  \bibnamefont{and} \bibinfo{author}{\bibfnamefont{L.~F.}
  \bibnamefont{Santos}}, 
    \bibinfo{journal}{Phys. Rev. A} \textbf{\bibinfo{volume}{89}},
  \bibinfo{pages}{043620} (\bibinfo{year}{2014}).

\bibitem[{\citenamefont{Torres-Herrera et~al.}()\citenamefont{Torres-Herrera,
  Vyas, and Santos}}]{Torres2014NJP}
\bibinfo{author}{\bibfnamefont{E.~J.} \bibnamefont{Torres-Herrera}},
  \bibinfo{author}{\bibfnamefont{M.}~\bibnamefont{Vyas}}, \bibnamefont{and}
  \bibinfo{author}{\bibfnamefont{L.~F.} \bibnamefont{Santos}},
  \bibinfo{journal}{New J. Phys.} \textbf{\bibinfo{volume}{16}},
  \bibinfo{pages}{063010} (\bibinfo{year}{2014}).
  
\bibitem[{\citenamefont{Genway et~al.}(2000)\citenamefont{Genway,
  Ho, Lee}}]{GenwayPRL2010}
\bibinfo{author}{\bibfnamefont{S.}~\bibnamefont{Genway}},
  \bibinfo{author}{\bibfnamefont{A.~F.}\bibnamefont{Ho}}, \bibnamefont{and}
  \bibinfo{author}{\bibfnamefont{D.~K.~K.} \bibnamefont{Lee}},
  \bibinfo{journal}{Phys. Rev. Lett.} \textbf{\bibinfo{volume}{105}},
  \bibinfo{pages}{260402} (\bibinfo{year}{2010}).

\bibitem[{\citenamefont{Sologubenko et~al.}(2000)\citenamefont{Sologubenko,
  Felder, Giann\`o, Ott, Vietkine, and Revcolevschi}}]{Sologubenko2000PRB}
\bibinfo{author}{\bibfnamefont{A.~V.} \bibnamefont{Sologubenko}},
  \bibinfo{author}{\bibfnamefont{E.}~\bibnamefont{Felder}},
  \bibinfo{author}{\bibfnamefont{K.}~\bibnamefont{Giann\`o}},
  \bibinfo{author}{\bibfnamefont{H.~R.} \bibnamefont{Ott}},
  \bibinfo{author}{\bibfnamefont{A.}~\bibnamefont{Vietkine}}, \bibnamefont{and}
  \bibinfo{author}{\bibfnamefont{A.}~\bibnamefont{Revcolevschi}},
  \bibinfo{journal}{Phys. Rev. B} \textbf{\bibinfo{volume}{62}},
  \bibinfo{pages}{R6108} (\bibinfo{year}{2000}).

\bibitem[{\citenamefont{Hess}(2007)}]{Hess2007}
\bibinfo{author}{\bibfnamefont{C.}~\bibnamefont{Hess}}, \bibinfo{journal}{Eur.
  Phys. J. Special Topics} \textbf{\bibinfo{volume}{151}}, \bibinfo{pages}{73}
  (\bibinfo{year}{2007}).

\bibitem[{\citenamefont{Hlubek et~al.}(2010)\citenamefont{Hlubek, Ribeiro,
  Saint-Martin, Revcolevschi, Roth, Behr, B\"uchner, and Hess}}]{Hlubek2010}
\bibinfo{author}{\bibfnamefont{N.}~\bibnamefont{Hlubek}},
  \bibinfo{author}{\bibfnamefont{P.}~\bibnamefont{Ribeiro}},
  \bibinfo{author}{\bibfnamefont{R.}~\bibnamefont{Saint-Martin}},
  \bibinfo{author}{\bibfnamefont{A.}~\bibnamefont{Revcolevschi}},
  \bibinfo{author}{\bibfnamefont{G.}~\bibnamefont{Roth}},
  \bibinfo{author}{\bibfnamefont{G.}~\bibnamefont{Behr}},
  \bibinfo{author}{\bibfnamefont{B.}~\bibnamefont{B\"uchner}},
  \bibnamefont{and} \bibinfo{author}{\bibfnamefont{C.}~\bibnamefont{Hess}},
  \bibinfo{journal}{Phys. Rev. B} \textbf{\bibinfo{volume}{81}},
  \bibinfo{pages}{020405(R) 1} (\bibinfo{year}{2010}).

\bibitem[{\citenamefont{Santos}(2009)}]{Santos2009JMP}
\bibinfo{author}{\bibfnamefont{L.~F.} \bibnamefont{Santos}},
  \bibinfo{journal}{J. Math. Phys} \textbf{\bibinfo{volume}{50}},
  \bibinfo{pages}{095211 (1} (\bibinfo{year}{2009}).

\bibitem[{\citenamefont{Gubin and Santos}(2012)}]{Gubin2012}
\bibinfo{author}{\bibfnamefont{A.}~\bibnamefont{Gubin}} \bibnamefont{and}
  \bibinfo{author}{\bibfnamefont{L.~F.} \bibnamefont{Santos}},
  \bibinfo{journal}{Am. J. Phys.} \textbf{\bibinfo{volume}{80}},
  \bibinfo{pages}{246} (\bibinfo{year}{2012}).

\bibitem[{\citenamefont{Zangara et~al.}(2013)\citenamefont{Zangara, Dente,
  Torres-Herrera, Pastawski, Iucci, and Santos}}]{Zangara2013}
\bibinfo{author}{\bibfnamefont{P.~R.} \bibnamefont{Zangara}},
  \bibinfo{author}{\bibfnamefont{A.~D.} \bibnamefont{Dente}},
  \bibinfo{author}{\bibfnamefont{E.~J.} \bibnamefont{Torres-Herrera}},
  \bibinfo{author}{\bibfnamefont{H.~M.} \bibnamefont{Pastawski}},
  \bibinfo{author}{\bibfnamefont{A.}~\bibnamefont{Iucci}}, \bibnamefont{and}
  \bibinfo{author}{\bibfnamefont{L.~F.} \bibnamefont{Santos}},
  \bibinfo{journal}{Phys. Rev. E} \textbf{\bibinfo{volume}{88}},
  \bibinfo{pages}{032913} (\bibinfo{year}{2013}).

\bibitem[{\citenamefont{Alcaraz et~al.}(1987)\citenamefont{Alcaraz, Barber,
  Batchelor, Baxter, and Quispel}}]{Alcaraz1987}
\bibinfo{author}{\bibfnamefont{F.~C.} \bibnamefont{Alcaraz}},
  \bibinfo{author}{\bibfnamefont{M.~N.} \bibnamefont{Barber}},
  \bibinfo{author}{\bibfnamefont{M.~T.} \bibnamefont{Batchelor}},
  \bibinfo{author}{\bibfnamefont{R.~J.} \bibnamefont{Baxter}},
  \bibnamefont{and} \bibinfo{author}{\bibfnamefont{G.~R.~W.}
  \bibnamefont{Quispel}}, \bibinfo{journal}{J. Phys. A}
  \textbf{\bibinfo{volume}{20}}, \bibinfo{pages}{6397} (\bibinfo{year}{1987}).

\bibitem[{\citenamefont{Jordan and Wigner}(1928)}]{Jordan1928}
\bibinfo{author}{\bibfnamefont{P.}~\bibnamefont{Jordan}} \bibnamefont{and}
  \bibinfo{author}{\bibfnamefont{E.}~\bibnamefont{Wigner}},
  \bibinfo{journal}{Z. Phys.} \textbf{\bibinfo{volume}{47}},
  \bibinfo{pages}{631} (\bibinfo{year}{1928}).

\bibitem[{\citenamefont{Bethe}(1931)}]{Bethe1931}
\bibinfo{author}{\bibfnamefont{H.~A.} \bibnamefont{Bethe}},
  \bibinfo{journal}{Z. Phys.} \textbf{\bibinfo{volume}{71}},
  \bibinfo{pages}{205} (\bibinfo{year}{1931}).

\bibitem[{\citenamefont{Kudo and Deguchi}(2004)}]{Kudo2004}
\bibinfo{author}{\bibfnamefont{K.}~\bibnamefont{Kudo}} \bibnamefont{and}
  \bibinfo{author}{\bibfnamefont{T.}~\bibnamefont{Deguchi}},
  \bibinfo{journal}{Phys. Rev. B} \textbf{\bibinfo{volume}{69}},
  \bibinfo{pages}{132404} (\bibinfo{year}{2004}).

\bibitem[{\citenamefont{Kudo and Deguchi}(2005)}]{Kudo2005}
\bibinfo{author}{\bibfnamefont{K.}~\bibnamefont{Kudo}} \bibnamefont{and}
  \bibinfo{author}{\bibfnamefont{T.}~\bibnamefont{Deguchi}},
  \bibinfo{journal}{J. Phys. Soc. Jpn.} \textbf{\bibinfo{volume}{74}},
  \bibinfo{pages}{1992} (\bibinfo{year}{2005}).

\bibitem[{\citenamefont{Brody et~al.}(1981)\citenamefont{Brody, Flores, French,
  Mello, Pandey, and Wong}}]{Brody1981}
\bibinfo{author}{\bibfnamefont{T.~A.} \bibnamefont{Brody}},
  \bibinfo{author}{\bibfnamefont{J.}~\bibnamefont{Flores}},
  \bibinfo{author}{\bibfnamefont{J.~B.} \bibnamefont{French}},
  \bibinfo{author}{\bibfnamefont{P.~A.} \bibnamefont{Mello}},
  \bibinfo{author}{\bibfnamefont{A.}~\bibnamefont{Pandey}}, \bibnamefont{and}
  \bibinfo{author}{\bibfnamefont{S.~S.~M.} \bibnamefont{Wong}},
  \bibinfo{journal}{Rev. Mod. Phys} \textbf{\bibinfo{volume}{53}},
  \bibinfo{pages}{385} (\bibinfo{year}{1981}).

\bibitem[{not({\natexlab{a}})}]{noteXXZ}
\bibinfo{note}{This is to be contrasted with the behavior of $M_n$ for the
  integrable $XXZ$ model, which is far from smooth~ \cite{Santos2012PRE}.}

\bibitem[{\citenamefont{Guhr et~al.}(1998)\citenamefont{Guhr,
  Mueller-Gr\"oeling, and Weidenm\"uller}}]{Guhr1998}
\bibinfo{author}{\bibfnamefont{T.}~\bibnamefont{Guhr}},
  \bibinfo{author}{\bibfnamefont{A.}~\bibnamefont{Mueller-Gr\"oeling}},
  \bibnamefont{and} \bibinfo{author}{\bibfnamefont{H.~A.}
  \bibnamefont{Weidenm\"uller}}, \bibinfo{journal}{Phys. Rep.}
  \textbf{\bibinfo{volume}{299}}, \bibinfo{pages}{189} (\bibinfo{year}{1998}).

\bibitem[{\citenamefont{Santos and Rigol}(2010{\natexlab{a}})}]{Santos2010PRE}
\bibinfo{author}{\bibfnamefont{L.~F.} \bibnamefont{Santos}} \bibnamefont{and}
  \bibinfo{author}{\bibfnamefont{M.}~\bibnamefont{Rigol}},
  \bibinfo{journal}{Phys. Rev. E} \textbf{\bibinfo{volume}{81}},
  \bibinfo{pages}{036206} (\bibinfo{year}{2010}{\natexlab{a}}).

\bibitem[{\citenamefont{Jacquod and Shepelyansky}(1997)}]{Jacquod1997}
\bibinfo{author}{\bibfnamefont{P.}~\bibnamefont{Jacquod}} \bibnamefont{and}
  \bibinfo{author}{\bibfnamefont{D.~L.} \bibnamefont{Shepelyansky}},
  \bibinfo{journal}{Phys. Rev. Lett.} \textbf{\bibinfo{volume}{79}},
  \bibinfo{pages}{1837} (\bibinfo{year}{1997}).

\bibitem[{\citenamefont{Izrailev}(1990)}]{Izrailev1990}
\bibinfo{author}{\bibfnamefont{F.~M.} \bibnamefont{Izrailev}},
  \bibinfo{journal}{Phys. Rep.} \textbf{\bibinfo{volume}{196}},
  \bibinfo{pages}{299} (\bibinfo{year}{1990}).
  
\bibitem[{\citenamefont{Gorin et al.~}(2006)}]{Gorin2006}
\bibinfo{author}{\bibfnamefont{T.}~\bibnamefont{Gorin}}, \bibinfo{author}{\bibfnamefont{T.}~\bibnamefont{Prosen}}, \bibinfo{author}{\bibfnamefont{T.H.}~\bibnamefont{Seligman}} \bibnamefont{and}
  \bibinfo{author}{\bibfnamefont{M.}~\bibnamefont{Znidaric}},
  \bibinfo{journal}{Phys. Rep.} \textbf{\bibinfo{volume}{435}},
  \bibinfo{pages}{33} (\bibinfo{year}{2006}).

\bibitem[{\citenamefont{Jacquod and Petitjean}(2009)}]{Jacquod2009}
\bibinfo{author}{\bibfnamefont{P.}~\bibnamefont{Jacquod}} \bibnamefont{and}
  \bibinfo{author}{\bibfnamefont{C.}~ \bibnamefont{Petitjean}},
  \bibinfo{journal}{Advances in Physics} \textbf{\bibinfo{volume}{58}},
  \bibinfo{pages}{67} (\bibinfo{year}{2009}).    

\bibitem[{not({\natexlab{b}})}]{noteAditi}
\bibinfo{note}{In \cite{SchiroARXIV} it is claimed that an exponential decay of
  the Loschmidt echo can be achieved for a ground state if a sequence of a
  global and then a local quench is performed.}

\bibitem[{\citenamefont{Jensen and Shankar}(1985)}]{Jensen1985}
\bibinfo{author}{\bibfnamefont{R.~V.} \bibnamefont{Jensen}} \bibnamefont{and}
  \bibinfo{author}{\bibfnamefont{R.}~\bibnamefont{Shankar}},
  \bibinfo{journal}{Phys. Rev. Lett.} \textbf{\bibinfo{volume}{54}},
  \bibinfo{pages}{1879} (\bibinfo{year}{1985}).

\bibitem[{\citenamefont{Deutsch}(1991)}]{Deutsch1991}
\bibinfo{author}{\bibfnamefont{J.~M.} \bibnamefont{Deutsch}},
  \bibinfo{journal}{Phys. Rev. A} \textbf{\bibinfo{volume}{43}},
  \bibinfo{pages}{2046} (\bibinfo{year}{1991}).

\bibitem[{\citenamefont{Srednicki}(1994)}]{Srednicki1994}
\bibinfo{author}{\bibfnamefont{M.}~\bibnamefont{Srednicki}},
  \bibinfo{journal}{Phys. Rev. E} \textbf{\bibinfo{volume}{50}},
  \bibinfo{pages}{888} (\bibinfo{year}{1994}).

\bibitem[{\citenamefont{Horoi et~al.}(1995)\citenamefont{Horoi, Zelevinsky, and
  Brown}}]{Horoi1995}
\bibinfo{author}{\bibfnamefont{M.}~\bibnamefont{Horoi}},
  \bibinfo{author}{\bibfnamefont{V.}~\bibnamefont{Zelevinsky}},
  \bibnamefont{and} \bibinfo{author}{\bibfnamefont{B.~A.} \bibnamefont{Brown}},
  \bibinfo{journal}{Phys. Rev. Lett.} \textbf{\bibinfo{volume}{74}},
  \bibinfo{pages}{5194} (\bibinfo{year}{1995}).

\bibitem[{\citenamefont{Flambaum et~al.}(1996)\citenamefont{Flambaum, Izrailev,
  and Casati}}]{Flambaum1996b}
\bibinfo{author}{\bibfnamefont{V.~V.} \bibnamefont{Flambaum}},
  \bibinfo{author}{\bibfnamefont{F.~M.} \bibnamefont{Izrailev}},
  \bibnamefont{and} \bibinfo{author}{\bibfnamefont{G.}~\bibnamefont{Casati}},
  \bibinfo{journal}{Phys. Rev. E} \textbf{\bibinfo{volume}{54}},
  \bibinfo{pages}{2136} (\bibinfo{year}{1996}).

\bibitem[{\citenamefont{Flambaum and Izrailev}(1997)}]{Flambaum1997}
\bibinfo{author}{\bibfnamefont{V.~V.} \bibnamefont{Flambaum}} \bibnamefont{and}
  \bibinfo{author}{\bibfnamefont{F.~M.} \bibnamefont{Izrailev}},
  \bibinfo{journal}{Phys. Rev. E} \textbf{\bibinfo{volume}{56}},
  \bibinfo{pages}{5144} (\bibinfo{year}{1997}).

\bibitem[{\citenamefont{Borgonovi et~al.}(1998)\citenamefont{Borgonovi,
  Guarnieri, Izrailev, and Casati}}]{Borgonovi1998}
\bibinfo{author}{\bibfnamefont{F.}~\bibnamefont{Borgonovi}},
  \bibinfo{author}{\bibfnamefont{I.}~\bibnamefont{Guarnieri}},
  \bibinfo{author}{\bibfnamefont{F.~M.} \bibnamefont{Izrailev}},
  \bibnamefont{and} \bibinfo{author}{\bibfnamefont{G.}~\bibnamefont{Casati}},
  \bibinfo{journal}{Phys. Lett. A} \textbf{\bibinfo{volume}{247}},
  \bibinfo{pages}{140} (\bibinfo{year}{1998}).

\bibitem[{\citenamefont{Borgonovi and Izrailev}(2000)}]{Borgonovi2000}
\bibinfo{author}{\bibfnamefont{F.}~\bibnamefont{Borgonovi}} \bibnamefont{and}
  \bibinfo{author}{\bibfnamefont{F.~M.} \bibnamefont{Izrailev}},
  \bibinfo{journal}{Phys. Rev. E} \textbf{\bibinfo{volume}{62}},
  \bibinfo{pages}{6475} (\bibinfo{year}{2000}).

\bibitem[{I01()}]{I01}
\bibinfo{note}{F.~M. Izrailev, Physica Scripta {\bf T90}, 95 (2001).}

\bibitem[{\citenamefont{Kollath et~al.}(2007)\citenamefont{Kollath,
  L{\"a}uchli, and Altman}}]{Kollath2007}
\bibinfo{author}{\bibfnamefont{C.}~\bibnamefont{Kollath}},
  \bibinfo{author}{\bibfnamefont{A.~M.} \bibnamefont{L{\"a}uchli}},
  \bibnamefont{and} \bibinfo{author}{\bibfnamefont{E.}~\bibnamefont{Altman}},
  \bibinfo{journal}{Phys. Rev. Lett.} \textbf{\bibinfo{volume}{98}},
  \bibinfo{pages}{180601} (\bibinfo{year}{2007}).

\bibitem[{\citenamefont{Manmana et~al.}(2007)\citenamefont{Manmana, Wessel,
  Noack, and Muramatsu}}]{Manmana2007}
\bibinfo{author}{\bibfnamefont{S.~R.} \bibnamefont{Manmana}},
  \bibinfo{author}{\bibfnamefont{S.}~\bibnamefont{Wessel}},
  \bibinfo{author}{\bibfnamefont{R.~M.} \bibnamefont{Noack}}, \bibnamefont{and}
  \bibinfo{author}{\bibfnamefont{A.}~\bibnamefont{Muramatsu}},
  \bibinfo{journal}{Phys. Rev. Lett.} \textbf{\bibinfo{volume}{98}},
  \bibinfo{pages}{210405} (\bibinfo{year}{2007}).

\bibitem[{\citenamefont{Rigol et~al.}(2008)\citenamefont{Rigol, Dunjko, and
  Olshanii}}]{Rigol2008}
\bibinfo{author}{\bibfnamefont{M.}~\bibnamefont{Rigol}},
  \bibinfo{author}{\bibfnamefont{V.}~\bibnamefont{Dunjko}}, \bibnamefont{and}
  \bibinfo{author}{\bibfnamefont{M.}~\bibnamefont{Olshanii}},
  \bibinfo{journal}{Nature} \textbf{\bibinfo{volume}{452}},
  \bibinfo{pages}{854} (\bibinfo{year}{2008}).

\bibitem[{\citenamefont{Rigol}(2009{\natexlab{a}})}]{rigol09STATa}
\bibinfo{author}{\bibfnamefont{M.}~\bibnamefont{Rigol}},
  \bibinfo{journal}{Phys. Rev. Lett.} \textbf{\bibinfo{volume}{103}},
  \bibinfo{pages}{100403} (\bibinfo{year}{2009}{\natexlab{a}}).

\bibitem[{\citenamefont{Rigol}(2009{\natexlab{b}})}]{rigol09STATb}
\bibinfo{author}{\bibfnamefont{M.}~\bibnamefont{Rigol}},
  \bibinfo{journal}{Phys. Rev. A} \textbf{\bibinfo{volume}{80}},
  \bibinfo{pages}{053607} (\bibinfo{year}{2009}{\natexlab{b}}).

\bibitem[{\citenamefont{Polkovnikov}(2011)}]{Polkovnikov2011}
\bibinfo{author}{\bibfnamefont{A.}~\bibnamefont{Polkovnikov}},
  \bibinfo{journal}{Ann. Phys. (N.Y.)} \textbf{\bibinfo{volume}{326}},
  \bibinfo{pages}{486} (\bibinfo{year}{2011}).

\bibitem[{\citenamefont{Santos et~al.}(2011)\citenamefont{Santos, Polkovnikov,
  and Rigol}}]{Santos2011PRL}
\bibinfo{author}{\bibfnamefont{L.~F.} \bibnamefont{Santos}},
  \bibinfo{author}{\bibfnamefont{A.}~\bibnamefont{Polkovnikov}},
  \bibnamefont{and} \bibinfo{author}{\bibfnamefont{M.}~\bibnamefont{Rigol}},
  \bibinfo{journal}{Phys. Rev. Lett.} \textbf{\bibinfo{volume}{107}},
  \bibinfo{pages}{040601} (\bibinfo{year}{2011}).

\bibitem[{\citenamefont{Santos et~al.}(2012{\natexlab{c}})\citenamefont{Santos,
  Polkovnikov, and Rigol}}]{Santos2012PRER}
\bibinfo{author}{\bibfnamefont{L.~F.} \bibnamefont{Santos}},
  \bibinfo{author}{\bibfnamefont{A.}~\bibnamefont{Polkovnikov}},
  \bibnamefont{and} \bibinfo{author}{\bibfnamefont{M.}~\bibnamefont{Rigol}},
  \bibinfo{journal}{Phys. Rev. E} \textbf{\bibinfo{volume}{86}},
  \bibinfo{pages}{010102} (\bibinfo{year}{2012}{\natexlab{c}}).

\bibitem[{\citenamefont{Santos and Rigol}(2010{\natexlab{b}})}]{Santos2010PREb}
\bibinfo{author}{\bibfnamefont{L.~F.} \bibnamefont{Santos}} \bibnamefont{and}
  \bibinfo{author}{\bibfnamefont{M.}~\bibnamefont{Rigol}},
  \bibinfo{journal}{Phys. Rev. E} \textbf{\bibinfo{volume}{82}},
  \bibinfo{pages}{031130} (\bibinfo{year}{2010}{\natexlab{b}}).

\bibitem[{\citenamefont{Rigol and Santos}(2010)}]{RigolSantos2010}
\bibinfo{author}{\bibfnamefont{M.}~\bibnamefont{Rigol}} \bibnamefont{and}
  \bibinfo{author}{\bibfnamefont{L.~F.} \bibnamefont{Santos}},
  \bibinfo{journal}{Phys. Rev. A} \textbf{\bibinfo{volume}{82}},
  \bibinfo{pages}{011604(R)} (\bibinfo{year}{2010}).

\bibitem[{\citenamefont{He and Rigol}(2013)}]{He2013}
\bibinfo{author}{\bibfnamefont{K.}~\bibnamefont{He}} \bibnamefont{and}
  \bibinfo{author}{\bibfnamefont{M.}~\bibnamefont{Rigol}},
  \bibinfo{journal}{Phys. Rev. A} \textbf{\bibinfo{volume}{87}},
  \bibinfo{pages}{043615} (\bibinfo{year}{2013}).

\bibitem[{\citenamefont{Torres-Herrera and Santos}(2013)}]{Torres2013}
\bibinfo{author}{\bibfnamefont{E.~J.} \bibnamefont{Torres-Herrera}}
  \bibnamefont{and} \bibinfo{author}{\bibfnamefont{L.~F.}
  \bibnamefont{Santos}}, \bibinfo{journal}{Phys. Rev. E}
  \textbf{\bibinfo{volume}{88}}, \bibinfo{pages}{042121}
  (\bibinfo{year}{2013}).

\bibitem[{\citenamefont{Schir{\'o} and Mitra}()}]{SchiroARXIV}
\bibinfo{author}{\bibfnamefont{M.}~\bibnamefont{Schir{\'o}}} \bibnamefont{and}
  \bibinfo{author}{\bibfnamefont{A.}~\bibnamefont{Mitra}},
\bibinfo{journal}{Phys. Rev. Lett.}
  \textbf{\bibinfo{volume}{112}}, \bibinfo{pages}{246401}
  (\bibinfo{year}{2014}).

\end{thebibliography}
\end{document}